\documentclass[useAMS,usenatbib,onecolumn,usegraphicx]{mn2e}
\usepackage{amssymb}

\title[Non-exponential  growth in thin  Keplerian discs under  toroidally-dominated magnetic fields
]
{
Regimes of the non-exponential temporal growth in
thin
Keplerian discs under toroidally-dominated magnetic fields
}
\author[Yuri M. Shtemler, Michael Mond, and Edward Liverts]
{Yuri M. Shtemler$^{1}$\thanks{E-mail:
shtemler@bgu.ac.il; mond@bgu.ac.il; eliverts@bgu.ac.il},  Michael Mond$^{1}$, and  Edward  Liverts$^{1}$\\
$^{1}$Department of Mechanical Engineering,  Ben-Gurion
University of the Negev,  P.O. Box 653, Beer-Sheva 84105,
Israel
}
\begin{document}

\date{Accepted ---. Received ----; in original form ----}

\pagerange{\pageref{firstpage}--\pageref{lastpage}} \pubyear{}

\maketitle

\label{firstpage}

\begin{abstract}
The linear stability of thin vertically-isothermal density-stratified Keplerian discs  in toroidally-dominated magnetic fields is treated by asymptotic expansions in the small aspect ratio of the discs. The discs are found to be spectrally stable.
 The great variety of possible initial conditions  leads to three regimes of non-exponential growth of perturbations, which are classified according to different relative levels of the in-plane and axial perturbed velocities.
The first two regimes of instability are characterized by the decoupling of the magneto-sonic (MS) and inertia-Coriolis (IC)  modes, as well as by   algebraic temporal growth of the perturbations, which are driven by either MS or IC modes (hereafter MS - and IC -regimes of instability, respectively). The third, mixed IC-MS regime of non-exponential, non-algebraic  growth is due only to non-axisymmetric perturbations. The latter  regime   is characterized by high radial and azimuthal wavenumbers,
and  growth time of the order of tens of rotating periods.
The mixed IC-MS regime  most likely exhibits the maximal growth as compared with the IC- and MS- regimes.
 In the first two regimes of instability the compressible MS mode plays a principal role either as the driver of the growth or the driven growing mode, while the mixed IC-MS regime is described by the Bousinesq approximation for incompressible fluid. The latter is  obtained as a natural limit of the expansion scheme. The presence of magnetic field in the mixed IC- MS regime may drastically increase the growth rates of the perturbations as compared with the pure hydrodynamic system.
\end{abstract}

\section{Introduction.}
Instabilities in thin magnetized Keplerian discs are of principal interest in astrophysics. The numerous possible equilibrium magnetic configurations and initial levels of perturbations in real discs lead to a multiplicity of admissible regimes of instabilities, whose different aspects  are shortly discussed below.
Traditionally, the results of the analytical study of the magneto-rotational instability (MRI) in infinitely long cylinders [\cite{Velichov 1959};  \cite{Chandrasechar 1960}] are adopted in order to derive criteria for the normal-mode (spectral) stability in thin discs [\cite{Balbus and Hawley 1991}]. These analytical results have been supported by numerical simulations, which describe the evolution of small perturbations and their subsequent development to turbulence [e.g. \cite{Brandenburg et al. 1995}; \cite{Hawley et al. 1996}; \cite{Bodo et al. 2008}; \cite{Regev and Umurhan 2008}].

{\it Adiabatic and vertically-isothermal disc models.} Two equations of state, adiabatic and vertically-isothermal ones, are commonly used for thin disc modeling [see e. g. \cite{Balbus and Hawley 1991}]. These two models correspond to quite different descriptions of the disc geometry: in the adiabatic model the disc has finite horizontal boundaries, while the horizontal boundaries of the vertically isothermal discs with the diffused mass density  expand to infinity, and a  finite effective disc thickness is properly defined.
In the adiabatic disc model the hydrodynamic flow exhibits a relatively weak singularity of the basic functions  at the horizontal edges of the unmagnetized discs [SMRRU2010], which becomes to be much stronger for magnetic field perturbations in toroidally dominated magnetic fields.
The model of the vertically- isothermal disc
%
%
is a fair approximation for thin discs under current consideration. It is characterized by the divergence of hydrodynamic velocities at infinity.  Nonetheless, the perturbed mass flux, magnetic field and density vanish at infinity [see \cite{Shtemler et al. 2011}, hereafter SML2011 and the current study].

{\it Incompressible- and compressible-fluid models.} The assumption of incompressibility of the perturbed motion is frequently employed in order to avoid  acoustic modes thus simplifying the system [e.g. \cite{Balbus and Hawley 1991}, (1992), \cite{Coleman et al. 1995}, \cite{Curry and Pudritz 1996}]. Although in thin Keplerian discs the axial acoustic crossing time is of the order of the rotation period, in some of the cases incompressibility is a reasonable assumption. Indeed, for the  poloidally dominated magnetic field configuration the acoustic waves decouple from the Alfv\'en-Coriolis modes and, as was numerically established, converges to incompressible limit at least at sufficiently large plasma beta [see e.g. SML2011 and references therein]. For toroidally-dominated magnetic field, the current study demonstrates that the Boussinesq model for incompressible fluid is rigorously obtained in the limit of large plasma beta for the most dagenerous regime of instability [see Section 5 below]. However accounting for compressibility is crucial for other  regimes of instability in toroidally-dominated magnetic field [see the current analysis in Sections 3 and 4].

{\it The local plane-wave approximations and the  asymptotic expansions in small disc's aspect ratio.} The simplest local approximation is the freezing equilibrium model that analyzes the shearing instability by freezing the parameters of the equilibrium state in the Lagrangian cylindrical coordinates $\{ r,\theta,z\}$ [e.g. \cite{Balbus and Hawley 1991}, (1992), \cite{Papaloizou and Terquem 1997}]. Thin discs are thus modeled within the Boussinesq approximation for incompressible fluids ignoring the local structure everywhere apart from the rotation radial shear effect that is the only inhomogeneity which is taken into account. Adopting Lagrangian shearing coordinates with $\Theta=\theta-\Omega(r)t$ (where $\Omega(r)$  is the Keplerian rotation frequency), and applying the local approximation to the stability problem with respect to in general non-axisymmetric (spiral) perturbations allows solutions in the following form: $f'\sim A(t)\exp(i k_r r+i k \Theta+i k_z z)$. The effect of transition to the Lagrangian coordinates is reduced to replacement of a fixed radial wave number with a shearing time-dependent one,  $\Delta k_r(t)=k_r -k t d\Omega/d r$. The time dependence of the radial wavenumber for non-axisymmetric modes does not allow a normal mode analysis, since it is inconsistent with Laplace transform in time. The resulting problem for the perturbations is solved by numerical integration in time of the corresponding ordinary differential equations with time-dependent coefficients.

Summarizing the results of the freezing equilibrium modeling, axially-dominated magnetic field is unstable on dynamical (orbital) time of the disc rotation to both axisymmetric and non-axisymmetric  perturbations [e.g. \cite{Tagger  et al. 1992}]. Whereas axisymmetric Alfv\'en-Coriolis waves are exponentially unstable for a mixed toroidally-poloidal equilibrium magnetic configurations in high plasma beta [\cite{Balbus and Hawley 1991}, \cite{Tagger  et al. 1992}] found that non-axisymmetric (spiral) perturbations are non- exponentially  unstable for the poloidally-dominated magnetic field already at low plasma beta. Furthermore, \cite{Balbus and Hawley 1992} found that pure toroidal magnetic configurations are stable with respect to axisymmetric modes, but exhibit a non-exponential instability under non-axisymmetric perturbations, which occurs  due to the time dependence of the radial wave number, $\Delta k_r(t)$.
%

The  shearing-box model is closely related to the freezing equilibrium approximation and is developed to study the shearing instability [see \cite{Umurhan and Regev 2004}, \cite{Regev and Umurhan 2008} and references therein]. The shearing-box model can be derived as a local expansion of the governing equations in the cylindrical reference frame and thus be expressed in Cartesian coordinates. It is widely accepted that the local shearing box approximation qualitatively reflects most necessary physics, such as differential rotation, compressibility etc. Transient growth of non-axisymmetric modes in pure toroidial magnetic field is exhibited within the shearing box approximation [\cite{Brandenburg and Dintras 2006}]. Shearing-box modeling carried out for non-axisymmetric perturbations of a background pure toroidal magnetic field within a range of resistivity and viscosity, demonstrates instability of the linear [\cite{Papaloizou and Terquem 1997}] and fully nonlinear system [\cite{Fleming et al. 2000}, \cite{Simon and Hawley 2009}]. Notwithstanding its success in predicting linear stability, the shearing box model has been recently criticized as to its viability in describing the ensuing nonlinear dynamical evolution of the system [\cite{ Umurhan et al. 2006}].

Some results of the local modeling are still mainly qualitative due to ignoring the radial as well as axial structure of the disc. An approach alternative to the shearing box model introduced by \cite{Regev 1983} is based on asymptotic expansions in a small aspect ratio,  $\epsilon$, of thin  discs. This approach allows to effectively describe the spatial inhomogeneity of the discs that makes it free from the locality constrains in both axial and radial variables.   The thin disc approximation allows the adequate modeling of perturbations with the characteristic axial scale of the perturbations of the order of the disc thickness, in compliance with the commonly accepted view that turbulence is driven on the scales of the disc thickness. When true thin disc geometry is taken into account, both the growth rates and number of spectrally unstable MRI modes are greatly reduced and decreased with the disc aspect ratio [\cite{Coppi and Keyes 2003}, \cite{Liverts and Mond 2009}]. These results for MRI strongly depend on the value of the plasma beta parameter [SML2011]. Thus, at large plasma betas starting from  $\beta=4\pi P_*/B_*^2\sim 5$, the results of the boundary-value problem analysis tend to those obtained by plane-wave assumption in the axial direction. In that case, the discrete stability curve that describes the growth rate vs the effective axial wave number for thin disc is well approximated by the corresponding continuous curve in the local modeling in \cite{Balbus and Hawley 1991}. However, for beta values close to the minimal critical value the number of unstable modes is small and the disc stability properties significantly deviate from those predicted by the local model. Additionally, the thin disc approximation may be effectively combined with the local approximations discussed above under additional restrictions on the plasma beta value (see Section 5 in the present analysis of discs in toroidally-dominated magnetic fields at large plasma beta).

{\it Non-exponential instabilities.} The MRI is commonly thought as a proper source of the normal-mode (exponential) instability under either poloidally- or toroidally-dominated or mixed poloidal-toroidal magnetic fields. This assumption is supported by analytical results for infinitely long cylinders or using the relevant local modeling [\cite{Balbus and Hawley 1991}, (1992), \cite{Coleman et al. 1995}, \cite{Curry and Pudritz 1996}, \cite{Ogilvie and Pringle 1996}, \cite{Terquem and Papaloizou 1996}, \cite{Papaloizou and Terquem 1997}, \cite{Pessah and Psaltis 2005}, \cite{Brandenburg and Dintras 2006}, \cite{Begelman and Pringle 2007}]. For poloidally-dominated magnetic fields the results of local modeling are consistent with those in thin disc approximation and exhibit the normal-mode instability (MRI) [SML2011], while thin discs in toroidally-dominated magnetic fields are rather exponentially stable. This renders the system prone to non-exponential instability, which emerged as the dynamical response of  thin discs either embedded in the toroidally-dominated magnetic field or even free from magnetic field at all. In general, two kinds of the non-exponential instability should be distinguished: algebraically and non-algebraically growth in time. It has been demonstrated that algebraic instability is the primary source of pure hydrodynamic motion in thin compressible adiabatic discs. For instance, this is established both numerically [\cite{Umurhan et al. 2006} and \cite{Rebusco et al. 2009}] and explicitly by asymptotic expansions in small $\epsilon$  [\cite{Shtemler et al. 2010}, hereafter SMRRU2010, where it is identified as the resonant as well as non-resonant interaction between inertia-Coriolis and sound waves]. Recently [SML2011] this approach has been generalized for Keplerian vertically-isothermal discs embedded in toroidally-dominated magnetic fields for one of the possible regimes (IC regime, see Section 4 below) of the instability investigated in the present study. It has been noted that such algebraic instability can lead to fruitful and physically sound results in the context of the well-known scenario for sub-critical transition to turbulence [\cite{Schmidt and Henningson 2001}]. 

{\it Poloidally - and toroidally-dominated magnetic fields.} Although the magnetic field configuration in real discs is rather unknown, the  poloidally dominated magnetic equilibrium configuration is commonly accepted in the initial stage of the disc rotation. MRI in an equilibrium under such circumstances is generally thought as the instability that leads to sustainable turbulence in accretion discs [see \cite{Balbus and Hawley 1991}]. However, according to observations and numerical simulations, the toroidal component of the magnetic field generated due to the differential rotation may become comparable to or even dominate the poloidal one [see e.g. \cite{Brandenburg et al. 1995}, \cite{Hawley et al. 1996}, \cite{Stone et al. 1996}, \cite{Hawley and Krolik 2002}, \cite{Proga 2003}]. This attracts attention to possible instabilities of Keplerian discs with a toroidally-dominated magnetic field.
For axisymmetric perturbations, the existence of two regimes of instability distinguished by equilibrium magnetic  configuration was demonstrated in thin disc approximation, namely either poloidally- or toroidally-dominated magnetic fields [SML2011]. In poloidally-dominated magnetic fields MRI (that is an axisymmetric normal-mode) was commonly found to be responsible for the most dangerous unstable mode, while under toroidally-dominated magnetic fields the non-exponential and non-axisymmetric governing modes of instability has been established in the current study. In SML2011 the toroidally-dominated magnetic configuration was treated for one of three admissible regimes considered in the current study: the non-exponential instability driven by either magnetosonic (MS) - or inertia-Coriolis (IC) -modes or  mixed IC-MS regime.

{\it Main goals, assumptions and structure of the current paper.}  The current work is aimed to further elucidate the physical mechanisms responsible for the generation of the most dangerous modes in compressible density-stratified Keplerian discs in toroidal-dominated magnetic fields taking into account the true thin-disc geometry. Thus, the present study  generalizes the stability analysis by SML2011  to non-axisymmetric perturbations and  extends it  to all possible regimes of instabilities.
The analysis is carried out under the basic assumptions that a toroidally-dominated magnetic field is settled during the time preceding to the initial time in the stability analysis, and the disc instability operates on several orbital periods and on wavelengths of the order of or less than the effective disc thickness determined by the fluid compressibility (as the ratio of the sound velocity to the Keplerian frequency).

The paper is organized as follows. The physical model for thin Keplerian discs that includes the dimensionless governing equations and their approximation to leading order in small aspect ratio for the steady-state disc are presented in the next Section. The governing equations and the results of analysis for perturbed thin discs in MS-regime of instability are presented in Section 3. Section 4 describes  similar results for IC-regime of instability. Section 5 presents the stability analysis for the mixed IC- MS regime. The role of magnetic fields on the perturbation growth for three regimes is estimated in Section 6 by setting both equilibrium and perturbed magnetic fields to zero, and consideration of the pure hydrodynamic limit. Summary and discussion are presented in Section 7.

\section{THE PHYSICAL MODEL FOR THIN KEPLERIAN DISCS.}
The stability of radially and axially-stratified rotating plasmas in thin vertically-isothermal discs is considered.  The discs are threaded by an equilibrium magnetic field whose toroidal component is larger than the other two components in a sense that will be defined below  (the poloidally-dominated magnetic field is considered in details in SML2011). Viscosity, electrical resistivity, and radiation effects are ignored.

\subsection{Governing equations.}
The dimensionless dynamical equations for vertically isothermal discs are the same as in SMRRU2010 and SML2011, and we  briefly recall them here:
\begin{equation}
\frac{ D \bf {V} }{Dt} = -\frac{ 1}{M_S^2}\frac{\nabla P}{n}-  \nabla \Phi+\frac{1}{\beta M^2_S} \frac{\bf{j}
\times {\bf{B}}}{n},\\
\label{1}
\end{equation}
\begin{equation}
\frac{\partial n} {\partial t}
               + \nabla \cdot(n {\bf{V}} ) =0,\\
\label{2}
\end{equation}
\begin{equation}
\frac{\partial {\bf{B} }} {\partial t}+
 \nabla \times
 {\bf{E}}=0,\,\nabla \cdot {\bf{B}}=0,
\label{3}
\end{equation}
\begin{equation}
 {\bf {E} }=  -  {\bf{V}} \times {\bf{B}},
\label{4}
\end{equation}
\begin{equation}
P =nT.
\label{5}
\end{equation}
Here $\nabla P=\bar{c}^2_S\nabla n$  for vertically isothermal discs, and the dimensionless equilibrium sound
speed is given by $\bar{c}^2_S =\partial P/\partial n\equiv T(r)$. Standard cylindrical coordinates
$\{r,\theta,z\}$   are adopted throughout the paper; $\bf{V}$  is the plasma velocity; $t$ is time; $D/Dt=\partial/\partial
t+(\bf{V}\cdot\nabla)$  is the material derivative; $\Phi(r,z)=-(r^2+z^2)^{-1/2}$  is the
 gravitational potential due to the central body; $\bf{B}$, $\bf{j}=\nabla\times\bf{B}$, and $\bf{E}$ are the magnetic field,    current density and
  electric field, respectivelly; $P=P_e+P_i$  is the total plasma pressure; $P_l=n_lT_l$  are the partial species
  pressures ($l=e,i$); $T=T_e=T_i$  is the plasma temperature; subscripts $e$ and $i$ denote electrons
  and ions, respectively. The
  positive direction of the $z$ axis is chosen according to positive Keplerian rotation. The dimensionless
  coefficients $M_S$  and $\beta$  are the Mach number and the characteristic plasma beta, respectively:
\begin{equation}
M_S=\frac{V_*} {c_{S*}},\,\, \beta=4\pi\frac{P_*} {B_{*}^2}.
\label{6}
\end{equation}
The asterisk denotes the corresponding dimensional characteristic values defined along with the dimensionless variables in SML2011. In particular, $V_*$, $c_{S*}$, $P_*$ and $B_*$
 are the characteristic velocity, sound speed, pressure and magnetic field, respectively. The characteristic radius value, $r_*$, the disc semi-thickness, $H(r_*)$,
 the inverse Keplerian frequency, $t_*=\Omega^{-1}(r_*)$, and the background toroidal magnetic field, $B_*=B_\theta(r_*)$, are taken as the characteristic dimensional scales for radial and axial lengths, time and magnetic field, and from the basis for estimations of the variables for all admissible regimes under consideration (see Tables 1 and 2 below). Additionally,  vanishing boundary conditions at infinity, $z=\pm \infty$,   are adopted for the magnetic field and density perturbed about an equilibrium state.

A common property of thin Keplerian discs is their highly
compressible motion with large Mach numbers [\cite{Frank et al.
2002}]. Furthermore,  the disc aspect ratio $\epsilon$  is a small parameter that is equal to the inverse Mach number:
\begin{equation}
\frac{1}{M_S}=\epsilon=\frac{H_*}{r_*}
\ll 1.
\label{7}
\end{equation}
The smallness of $\epsilon$  means that dimensionless axial coordinate is also small, i.e. $z/r_*\sim
\epsilon\,\,({\mid} z{\mid}  ^{<}_\sim H_*)$, and consequently the following rescaled quantities may be introduced in
order to
further apply the asymptotic expansions in $\epsilon $  [similar to \cite{Shtemler et al. 2009}; SMRRU2010; SML2011]:
\begin{equation}
\zeta=\frac{z}{\epsilon}\sim\epsilon^0,\,\,\,\bar{H}(r) =\frac{H(r)}{\epsilon}
\equiv \frac{\bar{c}_S(r)}{\bar{\Omega}(r)}\sim\epsilon^0,
\label{8}
\end{equation}
where $\bar{H}(r)$  is the scaled semi-thickness of the disc; $\bar{H}(r_*)=1$.

\subsection{Steady-state equilibrium configurations.}
It is first noted that the asymptotic expansion
for the time-independent gravitational potential is given by:
\begin{equation}
\Phi(r,\zeta)=\bar{\Phi}(r)+\epsilon^2\bar{\phi}(r,\zeta),\,\,\,
\bar{\Phi} (r)=-\frac{1}{r},\,\,\, \bar{\phi}(r,\zeta)
=\frac{1}{2}\zeta^2\bar{\Omega}^2(r)+O(\epsilon^2),\,\,\, r^{>}_\sim 1\gg\epsilon.
\label{9}
\end{equation}
Substituting (\ref{9}) into  (\ref{1})-(\ref{5}) and  setting  to zero the partial derivatives with respect to time yield
to leading order in $\epsilon$
 \begin{equation}
\frac{\bar{V}_\theta^2}{r}=\frac{d\bar{\Phi}(r)}{dr},\,\,\,\,
\frac{\bar{c}_S^2(r)}{\bar{n}}\frac{\partial \bar{n}}{\partial \zeta}=
-\frac{\partial \bar{\phi}(r,\zeta)}{ \partial
\zeta}.
\label{10}
\end{equation}
Thus, the solution of (\ref{10}) is given by:
\begin{equation}
V_\theta \cong\epsilon ^0\bar{V}_\theta(r)+O(\epsilon ^2),\,\,\,
 n\cong\epsilon ^0 \bar{n}= \epsilon ^0\bar{N}(r)\bar{\nu}(\eta),
\label{11}
\end{equation}
 and
\begin{equation}
\bar{V}_\theta(r)=r\bar{\Omega} (r),\,\,\, \bar{\Omega} (r)=r^{-3/2},\,\,\,
\bar{\nu}( \eta )=\exp(-\eta ^2/2), \ \ \ \eta=\zeta/\bar{H}(r),
\label{12}
\end{equation}
while  other components of the steady-state velocity are of lower order in $\epsilon$:   $V_r=o(\epsilon)$ and   $V_z=o(\epsilon^2)$  [SML2011].

The equilibrium magnetic configuration considered below is characterized by  scaling of the physical variables with  $\epsilon$  which generally may be written as $f(r,\zeta)= \epsilon^{\bar{S}} \bar{f}(r,\zeta)$. The magnetic field is assumed for simplicity to depend on the radius $r$ only. It is also assumed that the equilibrium is characterized by a toroidally-dominated equilibrium magnetic field that is of the order   $\epsilon^0$, while the axial component of the magnetic field is of  order   $\epsilon$. These assumptions determine  order in   $\epsilon$ of the rest of the physical variables [SML2011]:
 \begin{equation}
 B_r \cong o(\epsilon^2),\,\,\, B_\theta \cong\epsilon^0 \bar{B}_\theta(r),\,\,\, B_z \cong\epsilon  \bar{B}_z(r),
j_r \cong o(\epsilon^2),\,\,\,
j_\theta \cong\epsilon \bar{j}_\theta(r)= -\epsilon  \frac{d \bar{B}_z}{d r},\,\,\,
j_z \cong\epsilon^0 \bar{j}_z = \epsilon ^0 \frac{1}{r} \frac{d (r\bar{B}_\theta)}{d r}.\,\,\,
\label{13}
\end{equation}
All equilibrium variables are written in  leading order in  $\epsilon$, and depend on the radial variable only. The exceptions are the number density and the pressure that depend on the axial coordinate in a self-similar manner with radius-dependent amplitudes. In particular, the background toroidal and axial magnetic fields as well as the disc thickness and the amplitude factor, $\bar{N}(r)$, in the number density are arbitrary functions of the radial variable. Those functions specify the equilibrium state.
Each equilibrium variable is characterized by a gauge function  $\epsilon^{\bar{S}}$, where the parameters  $\bar{S}$  for the case of toroidally-dominated magnetic field are presented in Table 1 according to (\ref{10})-(\ref{13}).
 \begin{table*}
 \centering
 \begin{minipage}{140mm}
\caption{$\bar{S}$ for the equilibrium gauge functions in (\ref{14}).
}
\begin{tabular}{@{}cccccccc@{}}
& $n$ & $V_r$ & $V_\theta$ & $V_z$
& $B_r$ & $B_\theta$ & $B_z$
    \\
      \hline
       & $0$  &  $>2$  &   0  & $>2$
              &  $>2$  &   0  &  1
                    \\
  \end{tabular}
\end{minipage}
\end{table*}

\subsection{Gauge functions for the perturbations.}
In the general unsteady case the dependent variables are scaled in $\epsilon$  in the following way:
\begin{equation}
f(r,\zeta,t)=\epsilon^{\bar{S}}\bar{f}(r,\zeta)+\epsilon^{S'} f '(r,\zeta,t).
\label{14}
\end{equation}
Here $f$  stands for any dependent variable, the bar and the prime denote equilibrium and
 perturbed variables, respectively. The perturbed part of each variable is characterized by a gauge function $\epsilon^{S'}$ with some parameter  $S'$. As shown below, the relevant regimes of the instability in such systems are the IC-, MS- and mixed IC- MS regimes, named so according to the mode driving the perturbation growth. The gauge functions for all regimes are summarized in Table 2 ($\varkappa_r$, $\varkappa_\theta$ and $\varkappa_z$   are the characteristic length scales of the perturbations in the  radial, azimuthal and axial directions).
 \begin{table*}
 \centering
 \begin{minipage}{140mm}
\caption{$S'$ for the perturbed gauge functions in (\ref{14})  and for the characteristic length scales of the perturbations in the  radial, azimuthal and axial directions. }
\begin{tabular}{@{}cccccccccc@{}}
& $n$ & $V_r$ & $V_\theta$ & $V_z$
& $B_r$ & $B_\theta$ & $B_z$
& $\varkappa_r,\varkappa_\theta$ & $\varkappa_z$
    \\
        \hline
      axisymmetric MS regimes
       & $0$ & $2$   &   $2$     & $1$
             & $2$   &   $0$     & $3$      & $0$       & $1$
        \\
         \hline
       non-axisymmetric MS regimes
       & $0$ & $2$   & $2$     & $1$
             & $2$   & 0    & 1     & $0$    & $1$
         \\
           \hline
        axisymmetric/non-axisymmetric IC regimes
       & $0$ & $0$   & 0    & $1$
             & $0$   & 0    & 1     & $0$    & 1
         \\
                    \hline
        non-axisymmetric mixed IC-MS regimes
       & $0$ & $1$   & 1    & $1$
             & $0$   & 0    & 0     & $1$    & 1
         \\
\end{tabular}
\end{minipage}
\end{table*}


\subsection{New dependent and independent variables. }
Before turning to the solution of the linearized stability system of equations it is noted that, guided by the steady-state solution and the special form of the dependence on the azimuthal variable, it is convenient to introduce the following new independent variables:
\begin{equation}
\tau=\bar{\Omega}(r)t,\,\,\, \Theta=\theta-\bar{\Omega}(r)t,\,\,\, \rho=\int_0^r\frac{d r}{\bar{H}(r)},\,\,\,\eta=\frac{\zeta}{\bar{H}(r)},\,\,\, (\bar{H}(r) =\frac{\bar{c}_S(r)}{
\bar{\Omega}(r)}).
\label{15}
\end{equation}
The derivatives in the new and old variables are related as follows:
$$
\frac{\partial }{\partial t} =
\bar{\Omega}(\rho)\big{(}\frac{\partial }{\partial \tau }
-\frac{\partial }{\partial \Theta}\big{)},\,\,\,
\frac{\partial }{\partial \theta}=\frac{\partial }{\partial \Theta},\,\,\,
\frac{\partial}{\partial \zeta} = \frac{1 }{ \bar{H}(\rho)}\frac{\partial }{\partial \eta },\,\,\,
\frac{\partial}{\partial r} =
 \frac{1 }{ \bar{H}(\rho)}\big{[}\frac{\partial}{\partial \rho}
 +\frac{d\ln\bar{\Omega}}{d \rho}\tau\big{(}\frac{\partial }{\partial \tau }
-\frac{\partial }{\partial \Theta}\big{)}
 -  \frac{d \ln\bar{H}}{d \rho}
\eta\frac{\partial }{\partial \eta }\big{]}.\,\,\,\,\,\,\,\,\,\,\,\,\,\,\,\,\,\,\,\,\,\,\,\,\,\,\,\,\,\,\,\,\,\,\,\,\,\,\,\,\,\,\,\,\,\,\,\,
\,\,\,\,\,\,\,\,\,\,\,\,\,\,\,\,\,\,\,\,\,\,\,\,\,\,\,\,\,\,\,\,\,\,\,\,\,\,\,\,\,\,\,\,\,\,\,\,
$$
Here we keep without confusion the previous notations for $\bar{\Omega}$  and  $\bar{H}$, which now depend on the new radial variable  $\rho$. It is also convenient to introduce the following new dependent variables scaled by radial variable dependent factors:
\begin{equation}
{\bf {v}}(\rho,\eta,\tau) = \frac{{\bf{V}}{'}}{\bar{c}_S(r)},\,\,\,
\nu(\rho,\eta,\tau) = \frac{n'}{\bar{N}(r)\bar{\nu}(\eta)},\,\,\,
{\bf {b}}(\rho,\eta,\tau) = \frac{{\bf{B}}'}{\bar{B}_\theta(r)}.
\label{16}
\end{equation}
In addition, the following functions of the new radial variable $\rho$  are defined:
$$
\bar{\beta}_\theta (\rho) =\beta\frac{\bar{N}(\rho) \bar{c}_S^2 (\rho) }{\bar{B}_\theta^2 (\rho) },\,\,\,\,\,\,\,
\bar{S}(\rho)=\frac{\bar{B}_\theta(\rho)}{\bar{B}_z(\rho)},\,\,\,
\bar{r}(\rho)=\frac{r}{\bar{H}(\rho)},\,\,\, (\bar{H}(\rho) =\frac{\bar{c}_S(\rho)}{\bar{\Omega}(\rho)}),\,\,\,\,\,\,\,\,\,\,\,\,\,\,\,\,\,\,\,\,\,\,\,\,
\,\,\,\,\,\,\,\,\,\,\,\,\,\,\,\,\,\,\,\,\,\,\,\,\,\,\,\,\,\,\,\,\,\,\,\,\,\,\,\,\,\,\,\,\,\,\,\,
\,\,\,\,\,\,\,\,\,\,\,\,\,\,\,\,\,\,\,\,\,\,\,\,\,\,\,\,\,\,\,\,\,\,\,\,\,\,\,\,\,\,\,\,\,\,\,\,
\,\,\,\,\,\,\,\,\,\,\,\,\,\,\,\,\,\,\,\,\,\,\,\,\,\,\,\,\,\,\,\,\,\,\,\,\,\,\,\,\,\,\,\,\,\,\,\,
$$
\begin{equation}
\bar{D}_H(\rho)=\frac{d \ln\bar{H}}{d \rho},\,\,\,
\bar{D}_\Omega(\rho)=\frac{d\ln\bar{\Omega}}{d \rho},\,\,\,
\bar{D}_N(\rho)=\frac{d\ln\bar{N}}{d \rho},\,\,\,
\bar{D}_{B_\theta}(\rho)=\frac{d\ln B_\theta^2}{d \rho},\,\,\,
\bar{D}_{B_z}(\rho)=\frac{d\ln B_z^2}{d \rho}.
\label{17}
\end{equation}
Since the disc thickness is widely undetermined, and the above equations are significantly simplified in the case of disc of constant thickness, this will be adopted everywhere below
\begin{equation}
H(\rho)\equiv 1,\,\,\,
\bar{D}_H(\rho)\equiv 0,\,\,\,
\rho\bar{D}_\Omega(\rho)\equiv -3/2,\,\,\,
\bar{r}(\rho)\equiv r \equiv \rho.
\label{18}
\end{equation}

\section{THE MS-REGIME OF INSTABILITY.}
\subsection{Governing equations for the MS-regime.}
We start by substituting the decomposition of the total disturbed variables (\ref{14}) into the MHD system (\ref{1}) - (\ref{5}), and linearizing the resulting equations about the steady-state equilibrium solution (\ref{10})-(\ref{13}). By using the estimations for the equilibrium variables and the perturbed variables for the MS-regime (see Tables 1 and 2, respectively), the linearized momentum and mass balance equations are given to leading order in  $\epsilon$ in the following form:
\begin{equation}
\frac{\partial V_r'}{\partial t}
+\bar{\Omega}(r)\frac{\partial V_r'}{\partial \theta}
-2\bar{\Omega}(r)V_\theta'
=-\bar{c}_S^2(r)\frac{\partial }{\partial r}\big{(}\frac{ n'}{\bar{n}}\big{)}
-\frac{ 1}{\beta\bar{n}(r,\zeta)}
\big{[}
\bar{B}_\theta(r)\frac{\partial B_\theta'}{\partial r}
+\frac{ 1}{r^2}\frac{d (r^2 \bar{B_\theta})}{d r}\big{(}B_\theta'
-\frac{ n'}{2\bar{n}}\big{)}
\big{]},
\label{19}
\end{equation}
\begin{equation}
\frac{\partial V_\theta' }{\partial t}
+\bar{\Omega}(r)\frac{\partial V_\theta'}{\partial \theta}
+\frac{1}{2}\bar{\Omega}( r)V_r'
=-\bar{c}_S^2(r)\frac{ 1}{r}\frac{\partial }{\partial \theta}\big{(}\frac{ n'}{\bar{n}}\big{)}
+\frac{ 1}{\beta\bar{n}(r,\zeta)}\bar{B}_z(r)\frac{\partial B_\theta'}{\partial \zeta},
\label{20}
\end{equation}
\begin{equation}
\frac{\partial  V_z'}{\partial t}
+\bar{\Omega}(r)\frac{\partial V_z'}{\partial \theta}
=-\bar{c}_S^2(r) \frac{\partial}{\partial \zeta }\big{(}\frac{n'}{\bar{n}}\big{)}
-\frac{ 1}{\beta\bar{n}(r,\zeta)}
\bar{B}_\theta(r)\frac{\partial B_\theta'}{\partial \zeta},
\label{21}
\end{equation}
\begin{equation}
\frac{\partial n'}{\partial t}
+\bar{\Omega}(r)\frac{\partial n'}{\partial \theta}
+\frac{\partial (\bar{n}V_z')}{\partial \zeta }=0.
\label{22}
\end{equation}
 Maxwell's equations are now conveniently analyzed separately for the axisymmetric and non-axisymmetric regimes.

{\it Axisymmetric MS-regime.}
\begin{equation}
\frac{\partial B_r'}{\partial t}
-\bar{B}_z(r)\frac{\partial V_r'}{\partial \zeta}=0,
\label{23}
\end{equation}
\begin{equation}
\frac{\partial B_\theta'}{\partial t}
+\bar{B}_\theta(r)\frac{\partial  V_z'}{\partial \zeta}=0,
\label{24}
\end{equation}
\begin{equation}
\frac{\partial B_z'}{\partial t}
+\frac{ 1}{r}\frac{\partial [r \bar{B}_z(r)V_r']}{\partial r}=0.
\label{25}
\end{equation}
Note that the equations for both components of the perturbed poloidal magnetic field are decoupled from the rest of the system of equations, and may be dropped out from further consideration. In that case Eqs. (\ref{23}) and (\ref{25}) may be reduced to a single equation for the magnetic flux function, which guarantees the divergence-free condition  $\nabla \bold{B}'=0$:
\begin{equation}
 B_r'=-\frac{ 1}{r}\frac{\partial \Psi'}{\partial \zeta},\,\,\,\,\,
 B_z'=\frac{1}{r}\frac{\partial \Psi'}{\partial r}.\,\,\,\,\,
\label{26}
\end{equation}
Although this means that Eq. (\ref{23}) for $ B_r'$  can be dropped, and Eq. (\ref{25}) for $B_z'$  is sufficient for the further analysis, it will be convenient to calculate $ B_r'$   and   $ B_z'$  directly from Eqs. (\ref{23}) and (\ref{25}).

{\it Non-axisymmetric MS-regime.}
\begin{equation}
\frac{\partial B_r'}{\partial t}
+\bar{\Omega}(r)\frac{\partial B_r'}{\partial \theta}
-\bar{B}_\theta(r)\frac{ 1}{r}\frac{\partial V_r'}{\partial \theta}
-\bar{B}_z(r)\frac{\partial V_r'}{\partial \zeta}=0,
\label{27}
\end{equation}
\begin{equation}
\frac{\partial B_\theta'}{\partial t}
-r \bar{\Omega}(r)\frac{\partial B_z'}{\partial \zeta}
+\bar{B}_\theta(r)\frac{\partial  V_z'}{\partial \zeta}=0,
\label{28}
\end{equation}
\begin{equation}
\frac{\partial B_z'}{\partial t}
+\bar{\Omega}(r)\frac{\partial B_z'}{\partial \theta}
- \bar{B}_\theta(r)\frac{ 1}{r}\frac{\partial V_z'}{\partial \theta}=0.
\label{29}
\end{equation}
These equations for thin discs embedded in toroidally-dominated magnetic fields
($\bar{B}_r=0$, $\bar{B}_\theta\sim\epsilon^0$, $\bar{B}_z\sim\epsilon$) are quite different from their counterparts in the axisymmetric MC-regime. This is so not only due to the presence of terms with  $\partial /\partial \theta$, but also due to the appearance of $B_z'$  in the equation for  $B_\theta'$. Indeed, the  $\epsilon$-scaling of  $B_z'$   is different in those two cases:  $\epsilon B_z'\sim\epsilon$   in the non-axisymmetric case, while  $\epsilon B_z'\sim\epsilon^3$  in the axisymmetric case (see Table 2).

Applying the operator  $(\nabla\cdot)$ to the left-side of  (\ref{27}) - (\ref{29}) yields
\begin{equation}
\frac{\partial \nabla\cdot{\bf{B} }'} {\partial t}
-\bar{\Omega}(r)\frac{\partial \nabla\cdot{\bf{B} }'} {\partial \theta}
+ \bar{\Omega}(r)\nabla\cdot{\bf{B} }'=O(\epsilon^2),
\label{30}
\end{equation}
where
\begin{equation}
\nabla\cdot{\bf{B} }'
\equiv \frac{1}{r}\frac{\partial B'_\theta} {\partial \theta}
+ \frac{\partial B'_z} {\partial \zeta}
+O(\epsilon^2).
\label{31}
\end{equation}
Thus, the divergence-free condition,  $\nabla\cdot{\bf{B} }'=0$, satisfied at the initial instant, is guaranteed to be satisfied for all subsequent times in consequence of the Maxwell equations (\ref{27}) - (\ref{29}). Consequently, finding from the divergence-free condition
$\partial B'_z/\partial \zeta=-r^{-1}\partial B'_\theta/\partial \theta$, and substituting it into equation (\ref{28}) the latter may be rewritten in the standard form with the first two terms as material derivative of   $B'_\theta$:
\begin{equation}
\frac{\partial B_\theta'}{\partial t}
+\bar{\Omega}(r)\frac{\partial B_\theta'}{\partial \theta}
+\bar{\Omega}(r)\frac{\partial(r B_r')}{\partial r}
+\bar{B}_\theta(r)\frac{\partial  V_z'}{\partial \zeta}=0.
\label{32}
\end{equation}

It is first noted that the radial component of the perturbed magnetic field  $\epsilon^2 B'_r\sim \epsilon^2$ is small  compared with   other components of the perturbed magnetic field:
 $\epsilon^0 B'_\theta\sim \epsilon^0$ and  $\epsilon  B'_z\sim \epsilon$. Moreover, the input of $B'_z$  into the divergence-free condition $\nabla\cdot{\bf{B} }'=0$  is of  leading order in  $\epsilon$, while the corresponding input of $B'_r$  is negligibly small:
\begin{equation}
\epsilon^2\frac{1}{r}\frac{\partial(r B_r')}{\partial r}\ll
\epsilon^0\frac{1}{r}\frac{\partial B_\theta'}{\partial \theta}
\sim\epsilon^0\frac{\partial B_z'}{\partial \zeta}.
\label{33}
\end{equation}
Furthermore, $B_r'$  can be set equal to zero without loss of generality
\begin{equation}
 B_r'=0.
\label{34}
\end{equation}
Consequently Eq. (\ref{34}) substitutes  (\ref{27}) for non-axysimmetric perturbations, and the magnetic field is two-dimensional to leading order in  $\epsilon$, ${\bf{B} }'=\{0,\epsilon^2 B_\theta', \epsilon  B_z' \}$. As a result, the magnetic flux function $\Psi'$  (different from that introduced in (\ref{26}) for the axisymmetric case) can be used so that the divergence-free condition is identically satisfied:
\begin{equation}
 B_z'=\frac{ 1}{r}\frac{\partial \Psi'}{\partial \theta},\,\,\,\,\,
 B_\theta'=-\frac{\partial \Psi'}{\partial \zeta},\,\,\,\,\,
\label{35}
\end{equation}
and (\ref{29}) for $B_z'$  can be dropped from  further consideration. Again, however,  it is convenient to find  $B_z'$ and $B_\theta'$   directly from (\ref{29}) and (\ref{32}).

Examining the resulting system of equations, namely, (\ref{19})-(\ref{22}) and either (\ref{23})-(\ref{25}) (in the axisymmetric case) or (\ref{27})-(\ref{29}) (in the non-axisymmetric case), it is seen that the entire system decouples into three subsystems: (i) for the MS mode that consists of equations for the perturbed density,  $n'$, axial velocity,  $V_z'$, and toroidal magnetic field,  $B_\theta'$;  (ii) for the IC mode that consists of equations for the perturbed in-plane velocities $V_r'$  and  $B_\theta'$; and (iii) for the perturbed poloidal magnetic fields,  $B_r'$  and  $B_z'$. The systems of equations for the MS and IC modes have an invariant form for both axisymmetric and non-axisymmetric perturbations, as distinct from equations for perturbed poloidal magnetic field which should be described separately for axisymmetric and non-axisymmetric modes. Thus, the above mentioned subsystem (iii) for the perturbed poloidal magnetic fields can be dropped in the non-axisymmetric case, since   $B_r'=0$  and  $B_z'$   can be expressed through the magnetic flux function (\ref{35}) or, alternatively, through  $B_\theta'$, while in the axisymmetric case, introducing the magnetic flux function (\ref{26}) allows to reduce the subsystem to only one component of the perturbed poloidal magnetic field, e.g.  $B_z'$. Let us now rewrite the governing system for the MS mode in terms of the new variables (\ref{15})-(\ref{18}).

{\it The MS mode of the MS-regime.}
The equations for the in-plane components of the magnetic field decouple from the above system of governing equations, while the perturbed toroidal magnetic field,   $B_\theta'$, together with the equations for the perturbed   density and axial velocity, $n'$  and  $V_z'$, respectively, decouple from the equations for the IC mode, i.e. for $V_r'$  and   $V_\theta'$.

Thus, in the new variables (\ref{15})-(\ref{18}) the MS subsystem of equations yields for the perturbed axial velocity, density and toroidal magnetic field,  $\nu$, $v_z$  and  $b_\theta$:
\begin{equation}
\frac{\partial  v_z}{\partial \tau}
+\frac{\partial \nu}{\partial \eta }
+\frac{1}{\bar{\beta}_\theta(\rho)}\frac{1}{\bar{\nu}(\eta)}\frac{\partial b_\theta}{\partial \eta }=0,
\label{36}
\end{equation}
\begin{equation}
\frac{\partial  \nu}{\partial \tau}
+\frac{1}{\bar{\nu}(\eta)}\frac{\partial [\bar{\nu}(\eta) v_z]}{\partial \eta }=0,
\label{37}
\end{equation}
\begin{equation}
\frac{\partial b_\theta}{\partial \tau}
+\frac{\partial v_z}{\partial \eta }=0.
\label{38}
\end{equation}
Note that all equations for MS mode have an invariant form for both axisymmetric and non-axisymmetric modes, and are also free from  radial derivatives.

{\it The IC mode  of the MS-regime.}
Similarly, the following subsystem for perturbed in-plane velocities describes the IC modes:
\begin{equation}
 \frac{\partial v_r}{\partial \tau}
-2v_\theta=\tau\bar{D}_\Omega(\rho)
\big{\{}
\frac{\partial \nu}{\partial \Theta }
+\frac{1}{\bar{\beta}_\theta(\rho)}\frac{1}{\bar{\nu}(\eta)}\frac{\partial b_\theta }{\partial\Theta}
+\frac{1}{\bar{\nu}(\eta)}\frac{\partial [\bar{\nu}(\eta)v_z]}{\partial\eta}
\big{\}}
+\frac{1}{\bar{\beta}_\theta(\rho)}\frac{1}{\bar{\nu}(\eta)}
\frac{\partial v_z}{\partial\eta}
+\frac{1}{\bar{\beta}_\theta(\rho)}\frac{1}{\bar{\nu}(\eta)}L_0 b_\theta
+N_0\nu,
\label{39}
\end{equation}
\begin{equation}
\frac{\partial v_\theta }{\partial \tau}
+\frac{1}{2}v_r=-\frac{1}{\rho}\frac{\partial \nu}{\partial \Theta }
+\frac{1}{\bar{S}(\rho)\bar{\beta}_\theta(\rho)}\frac{1}{\bar{\nu}(\eta)}
\frac{\partial b_\theta }{\partial\eta},
\label{40}
\end{equation}
where
$$
L_0 b_\theta\equiv-\frac{\partial b_\theta }{\partial\rho}
-\big{[}
\frac{1 }{2}\bar{D}_{B\theta}(\rho)+\frac{2 }{\rho}
\big{]}b_\theta,\,\,\,\,\,\,\,\,
N_0\nu\equiv-\frac{\partial \nu }{\partial\rho}
+\frac{1 }{2}\frac{1}{\bar{\beta}_\theta(\rho)}\frac{1}{\bar{\nu}(\eta)}
\big{[}
\frac{1 }{2}\bar{D}_{B\theta}(\rho)+\frac{2 }{\rho}
\big{]}\nu\,\,\,\,\,\,\,\,\,\,\,\,\,\,\,\,\,\,\,\,\,\,\,\,\,\,\,\,\,\,\,
\,\,\,\,\,\,\,\,\,\,\,\,\,\,\,\,\,\,\,\,\,\,\,\,\,\,\,\,\,\,\,\,\,\,\,\,\,\,\,\,\,
\,\,\,\,\,\,\,\,\,
\,\,\,\,\,\,\,\,\,\,\,\,\,\,\,\,\,\,\,\,\,\,\,\,\,\,\,\,\,\,\,\,\,\,\,\,\,\,\,\,\,\,
$$
with  $\nu$, $v_z$  and $b_\theta$  in the right-hand sides of (\ref{39})- (\ref{40}) determined from (\ref{36})- (\ref{38}). As it is evident from (\ref{39}) - (\ref{40}), the MS mode acts as a driving force (hence the term MS regime. Note that the first  term in the right-hand side of (\ref{39}) is proportional to time. It arises from the radial derivatives of the perturbed density and toroidal magnetic field due to the radial shear of the Keplerian angular velocity (i.e.  $\bar{D}_\Omega(\rho)\neq0$).

{\it The poloidal magnetic field in the MS-regime.}
As  mentioned above the equations for the perturbed poloidal magnetic field are described separately for the axisymmetric and non-axisymmetric modes by Eqs. (\ref{23}), (\ref{25}) and (\ref{29}), (\ref{34}), respectively:

{\it (i) The axisymmetric case. }
\begin{equation}
 \frac{\partial b_r}{\partial \tau}=
\frac{1}{\bar{S}(\rho)}
\frac{\partial v_r}{\partial \eta },
\label{41}
\end{equation}
\begin{equation}
 \frac{\partial b_z}{\partial \tau}=
 - \tau^2
\frac{\bar{D}_\Omega^2(\rho)}{\bar{S}(\rho)}\frac{1}{\bar{\nu}(\eta)}
\frac{\partial [\bar{\nu}(\eta)v_z]}{\partial \eta }
- \tau\frac{\bar{D}_\Omega(\rho)}{\bar{S}(\rho)}\frac{1}{\bar{\nu}(\eta)}
\big{[}
\frac{1}{\bar{\beta}_\theta(\rho)}\frac{\partial v_z}{\partial \eta }
+2\bar{\nu}(\eta)v_\theta+\frac{1}{\bar{\beta}_\theta(\rho)}
L_0 b_\theta
+N_0\nu
\big{]}
-\frac{1}{\bar{S}(\rho)}M_0 v_r,
\label{42}
\end{equation}
where
$$
M_0 v_r\equiv
\frac{\partial v_r}{\partial \rho }
+\big{[}
\bar{D}_\Omega(\rho)+\frac{1}{2}\bar{D}_{Bz}(\rho)\big{]}v_r
+\frac{1}{ \rho }v_r.
\,\,\,\,\,\,\,\,\,\,\,\,\,\,\,\,\,\,\,\,\,\,\,\,\,\,\,\,\,\,\,\,\,\,\,\,\,\,\,\,\,\,\,\,\,\,\,\,
\,\,\,\,\,\,\,\,\,\,\,\,\,\,\,\,\,\,\,\,\,\,\,\,\,\,\,\,\,\,\,\,\,\,\,\,\,\,\,\,\,\,\,\,\,\,\,\,\,
\,\,\,\,\,\,\,\,\,\,\,\,\,\,\,\,\,\,\,\,\,\,\,\,\,\,\,\,\,\,\,\,\,\,\,\,
\,\,\,\,\,\,\,\,\,\,\,\,\,\,\,\,\,\,\,\,\,\,\,\,\,\,\,\,\,\,\,\,\,\,\,\,\,\,\,\,\,\,\,\,\,\,\,\,\,\,\,\,\,\,\,\,\,\,\,\,
\,\,\,\,\,\,\,\,\,\,\,\,\,\,\,\,\,\,\,\,\,\,\,\,\,\,\,\,\,\,\,\,\,\,\,\,
$$

{\it (ii) The non-axisymmetric case. }
\begin{equation}
 b_r=0,
\label{43}
\end{equation}
\begin{equation}
 \frac{\partial b_z}{\partial \tau}=
\frac{1}{ \rho } \frac{\partial v_z}{\partial \Theta}.
\label{44}
\end{equation}
Thus, for the axisymmetric modes the amplitudes of the perturbed poloidal magnetic field,  $b_z$, grows quadratically with time, while  in the non-axisymmetric case it is constant.

\subsection{ Normal-mode stability  for the MS-regime. }
We start by representing the perturbations as follows:
\begin{equation}
f(\rho,\Theta,\eta,\tau)=\exp[-i \lambda\tau+im\Theta ] \hat{f}(\rho,\eta)+c.c.\,\,\,
\label{45}
\end{equation}
Here $f$ stands for any perturbed variable,   $\lambda$ is the  eigenvalue, and  $m=0,\pm 1,\pm 2,... $ is the azimuthal wavenumber. Since the resulting equations which describe the temporal growth of the perturbations do not contain  radial derivatives, the normal-mode form (\ref{45}) for the  characteristic length scale  $\varkappa_r\sim\epsilon^0$ is used without assuming an exponential form in the radial direction.

{\it The MS mode of the MS-regime. }
Substituting the equilibrium number density $\bar{\nu}(\eta)=\exp(-\eta^2/2)$  and the ansatz (\ref{45}) into the subsystem (\ref{36}) - (\ref{38}) for the MS mode, the following single ordinar differential equation is obtained for  the toroidal magnetic field $\hat{b}_\theta$:
\begin{equation}
 \frac{d^2 \hat{b}_\theta}{d \eta^2 }
-\eta  \frac{\bar{\beta}_\theta \bar{\nu}(\eta)-1}
{\bar{\beta}_\theta \bar{\nu}(\eta)+1}\frac{d \hat{b}_\theta}{d \eta }
+(\lambda^2-2)
\frac{\bar{\beta}_\theta \bar{\nu}(\eta)}{\bar{\beta}_\theta \bar{\nu}(\eta)+1}
 \hat{b}_\theta
=0,\,\,\,\,\,\
 \hat{b}_\theta=0,\,\,\,\,\,\,\,\,\,\,\mbox{at}\,\,\,\,\,\eta=\pm\infty,
\label{46}
\end{equation}
while the axial velocity and density, $\hat{v}_z$  and  $\hat{\nu}$, can be easily expressed through  $\hat{b}_\theta$:
\begin{equation}
\hat{v}_z=i\lambda\int_\infty^\eta{\hat{b}_\theta}d\eta,\,\,\,\,\,\,\,\,\,\,\,\,
\hat{\nu}=\hat{b}_\theta-\eta\int_\infty^\eta{\hat{b}_\theta}d\eta.
\label{47}
\end{equation}
These equations for the MS mode in the MS-regime are similar to those in SML2011 for toroidally dominated magnetic fields. Summarizing those results we represent the approximate solution of (\ref{46}) obtained by applying the WKB approximation:
\begin{equation}
\hat{b}_\theta=Q(\eta)\exp(\int{\mu}d\eta),\,\,\,\,\,
\mu=\frac{\eta}{2} \frac{\bar{\beta}_\theta \bar{\nu}(\eta)-1}
{\bar{\beta}_\theta \bar{\nu}(\eta)+1}.
\label{48}
\end{equation}
The calculations are simplified if it is additionally assumed that $\bar{\beta}_\theta \gg 1$. This yields
\begin{equation}
Q=\frac{W}{\sqrt{\mid\varkappa(\eta)\mid}}\exp[\pm i\int \varkappa(\eta) d\eta]\,\,\,\,\,\,\,\mbox{and}\,\,\,\,\,\,\,
Q=\frac{W}{\sqrt{\mid\varkappa(\eta)\mid}}\exp[- \int \mid\varkappa(\eta)\mid d\eta].
\label{49}
\end{equation}	
Approximate expressions for  $\mu(\eta)$ and  $\varkappa(\eta)$  to leading order in  $\bar{\beta}_\theta $ are:
$$
 \mu (\eta)\approx \frac{1}{2}\eta,\,\,\,\,\,\,
 \varkappa^2(\eta)\approx -\frac{1}{4}(\eta^2-\eta^2_*),\,\,\,\,(\eta_*\approx 2\mid\lambda\mid).
 \,\,\,\, \,\,\,\, \,\,\,\, \,\,\,\, \,\,\,\, \,\,\,\, \,\,\,\, \,\,\,\, \,\,\,\,
\,\,\,\, \,\,\,\, \,\,\,\, \,\,\,\, \,\,\,\, \,\,\,\, \,\,\,\, \,\,\,\, \,\,\,\,
\,\,\,\, \,\,\,\, \,\,\,\, \,\,\,\, \,\,\,\, \,\,\,\, \,\,\,\, \,\,\,\, \,\,\,\,
\,\,\,\, \,\,\,\, \,\,\,\, \,\,\,\, \,\,\,\, \,\,\,\, \,\,\,\, \,\,\,\, \,\,\,\,
\,\,\,\, \,\,\,\, \,\,\,\, \,\,\,\, \,\,\,\, \,\,\,\, \,\,\,\, \,\,\,\, \,\,\,\,
$$
 Equations (\ref{49}) describe the oscillating solutions for  $\varkappa^2(\eta)>0$, $\mid\eta\mid<\eta_*$  and monotonically decreasing for  $\varkappa^2(\eta)<0$,  $\eta_*<\mid\eta\mid<\infty$, respectively, where the turning points $\eta=\pm\eta_*$   are the zeros of  $\varkappa(\eta)$.

To completely determine the  solution,  the matching conditions
betweenthe inner ($\mid \eta \mid < \eta_*$) and outer ($\mid \eta \mid > \eta_*$) regions  are applied in order to obtain the coefficients in (\ref{48})  and the eigenvalue equation. The result is the following Bohr-Zommerfeld condition [\cite{Landau and Lifshits 1997}] that determines the dispersion relation for the eigenvalues  $\lambda$
\begin{equation}
\int_{-\eta_*}^{+\eta_*} \varkappa(\eta)  d\eta=\pi(k+\frac{1}{2}),\,\,\,\
\mbox{and}\,\,\,\,\lambda =\lambda_{MS}=\pm \sqrt{k+1}\approx \pm \sqrt{k},
\label{50}
\end{equation}
where   $k$ is the number of zeros of the solution for $\hat{b}_\theta$, and  (\ref{50})
is valid for sufficiently large values of $\lambda$ and $k$, as  required in the WKB approximation.

{\it The IC mode of the MS-regime. }
 For the non-resonant case with non-equal frequencies of IC and MS modes, the perturbed axial velocity, torodial component of the magnetic field and density are zero, i.e.
\begin{equation}
\hat{v}_z=\hat{\nu}=\hat{b}_\theta=0.
\label{51}
\end{equation}
Therefore, the perturbed in-plane velocity is described by a set of equations that governs the dynamics of the IC modes. Substituting the ansatz (\ref{49}) and (\ref{42}) into (\ref{39}) - (\ref{40}) yields [SML2011]:
\begin{equation}
-i\lambda \hat{v}_r- 2\hat{v}_\theta=0,\,\,\,\,\,\,
-i\lambda \hat{v}_\theta+\frac{1}{2}\hat{v}_r =0.\,\,\,\,\,\,
\label{52}
\end{equation}
The corresponding dispersion relation is therefore given by
\begin{equation}
\lambda =\lambda_{IC}=\pm 1,\,\,\,\,\,
\label{53}
\end{equation}
which represents  stable epicyclical oscillations in the disc plane. As the axial and radial coordinates play the role of passive parameters, the eigenfunctions of the IC modes are determined up to arbitrary amplitude $A(\rho,\eta)$  from (\ref{52})-(\ref{53}):
\begin{equation}
 \hat{v}_\theta=\mp\frac{1}{2}i\hat{v}_r=\mp\frac{1}{2}iA(\rho,\eta),\,\,\,\,\,\,
 A(\rho,\eta)=F(\rho)G(\eta).
\label{54}
\end{equation}
Since any special form of the function  $A(\rho,\eta)$  represents some given set of initial conditions, the self-similar separable form of the planar velocities with the arbitrary functions $F(\rho)$  and  $G(\eta)$ is considered as an example.

\subsection{Algebraic temporal growth of perturbations in the MS -regime.}
A detailed account of resonant as well as  non-resonant  excitation that lead to quadratic and linear temporal growth, respectively, is
 given for pure hydrodynamic systems [SMRRU2010 and SML2011]. Here, however, the eigenvalues  $\lambda_{IC}=\pm 1$ of the IC normal modes in (\ref{53}) do not coincide with those for the MS modes in (\ref{50}),  $\lambda_{MS}\approx\pm \sqrt{k}$, in the large $k$ approximation adopted,  $\lambda_{MS}\neq\lambda_{IC}$,  the resonance condition is not satisfied. Focusing therefore on the IC modes that are non-resonantly driven by MS modes $\{\hat{b}_\theta,\hat{v}_z, \hat{\nu}\}$,
 and consequently on the linear temporal growth of the amplitude, the variables that characterize the IC-modes are given by:
\begin{equation}
\{v_r,v_\theta\}=\tau\{\hat{v}_r^{(1)},\hat{v}_\theta^{(1)}\}(\rho,\eta)
\exp(-i\lambda_{MS}\tau+im\Theta)+c.c.,\,\,\,\,\,\,
\{b_r,b_z\}=\{\tau^{j}\hat{b}_r^{(j)},\tau^{l}\hat{b}_z^{(l)}\}(\rho,\eta)
\exp(-i\lambda_{MS}\tau+im\Theta)+c.c.,\,\,\,\,\,\,
\label{55}
\end{equation}
where $m=0,\pm1,\pm2,...$  is the azimuthal wavenumber; $\hat{v}_r^{(1)}$,  $\hat{v}_\theta^{(1)}$ and $\hat{b}_r^{(j)}$,  $\hat{b}_z^{(l)}$    are amplitudes that depend on $\rho$ and  $\eta$, while the parameters $j$ and $l$ will be specified below; the radial magnetic field   is either zero for non-axisymmetric case or, as for axisymmetric case, can be found from Eq. (\ref{41}). Substituting (\ref{55}) into the system (\ref{39})-(\ref{40}) for $v_r$  and  $v_\theta$, and keeping only terms of the highest powers in $\tau$ (the rest of the terms in the solutions are dropped here for brevity),  yield for both axisymmetric and non-axisymmetric cases:
\begin{equation}
\frac{\hat{v}_r^{(1)}(\rho,\eta)}{i\lambda_{MS}}=
2\hat{v}_\theta^{(1)}(\rho,\eta)=\frac{1}{\lambda_{MS}^2-1}
\bar{D}_\Omega(\rho)\big{\{}im\big{[}\hat{\nu}+
\frac{1}{\bar{\beta}_\theta}\frac{1}{\bar{\nu}(\eta)}\hat{b}_\theta\big{]}
+\frac{1}{\bar{\nu}(\eta)}\frac{\partial[\bar{\nu}(\eta)\hat{v}_z]}{\partial\eta}
\big{\}},\,\,\,\,\,\,
\label{56}
\end{equation}
where  $\hat{\nu}$, $\hat{v}_z$ and  $\hat{b}_\theta$ are given by (47) and (48); the terms proportional to azimuthal wavenumber $m$ and shear coefficient $\bar{D}_\Omega(\rho)$  arise due to the joint non-axisymmetric and shear effects, while the rest of the terms proportional only to   $\bar{D}_\Omega(\rho)$  are entirely due to shear. Substituting the normal-mode ansatz (\ref{45}) into (\ref{41})-(\ref{42}) for  poloidal magnetic fields  $b_r$, $b_z$  in the axisymmetric case ($m=0$) and keeping only terms of the highest power in $\tau$  yield
\begin{equation}
\hat{b}_r^{(1)}(\rho,\eta)=\frac{1}{\bar{S}(\rho)}\frac{\partial\hat{v}_r^{(1)}}{\partial\eta},\,\,\,\,\,\,
\hat{b}_z^{(2)}(\rho,\eta)=\frac{1}{i\lambda_{MS}}
\frac{\bar{D}_\Omega^2(\rho)}{\bar{S}(\rho)}\frac{1}{\bar{\nu}(\eta)}
\frac{[\partial\bar{\nu}(\eta)\hat{v}_z]}{\partial\eta}.
\label{57}
\end{equation}
Similarly, (\ref{43}) and (\ref{44}) for $b_r$, $b_z$  in the non-axisymmetric case ($m\neq 0$) yield
\begin{equation}
\hat{b}_r^{(0)}(\rho,\eta)=0,\,\,\,\,
\hat{b}_z^{(0)}(\rho,\eta)=\frac{m}{\lambda_{MS}}
\frac{1}{\rho}\hat{v}_z.
\label{58}
\end{equation}
Thus, for  both axisymmetric and non-axisymmetric cases the equilbrium toroidal magnetic field gives rise to algebraic instability that is driven by the MS modes.
In the axisymmetric case the in-plane perturbed velocities and the perturbed radial magnetic field grow linearly in time, while the perturbed axial magnetic field grows quadraticaly in time. In turn, in the non-axisymmetric case there is a linear temporal growth of the in-plane perturbed velocities without algebraic growth in the perturbed magnetic field. The interesting fact is that the non-axisymmetric modes of the MS-regime of instability do not go over to the axisymmetric value at small azimuthal wavenumbers. This occurs due to nonuniform behavior of the perturbed axial magnetic field in $\epsilon$  that is  $\sim\epsilon$  and  $\sim\epsilon^3$  in the non-axisymmetric and axisymmetric cases, respectively.

\section{THE IC-REGIME OF INSTABILITY.}

\subsection{Governing equations for the IC -regime. }
We again start by linearizing the MHD Eqs. (\ref{1}) - (\ref{6}) about the steady-state equilibrium solution (\ref{10})-(\ref{13}) and substituting the appropriate expansion of the total disturbed variables (\ref{14}) for the equilibrium and perturbed variables for the IC-regime (see Tables 1 and 2, respectively). Keeping  terms of leading order in $\epsilon$  the system of equations is decoupled into the following  subsystems reformulated, as for the MS-regime, in terms of the variables defined in (\ref{15}) - (\ref{18}):

{\it The IC mode of the IC-regime. }
\begin{equation}
 \frac{\partial v_r}{\partial \tau}
-2v_\theta=0,
\label{59}
\end{equation}
\begin{equation}
\frac{\partial v_\theta }{\partial \tau}
+\frac{1}{2}v_r=0.
\label{60}
\end{equation}

{\it The  MS mode of the IC-regime. }
\begin{equation}
\frac{\partial  v_z}{\partial \tau}
+\frac{\partial \nu}{\partial \eta }
+\frac{1}{\bar{\beta}_\theta(\rho)}\frac{1}{\bar{\nu}(\eta)}\frac{\partial b_\theta}{\partial \eta }=0,
\label{61}
\end{equation}
\begin{equation}
\frac{\partial  \nu}{\partial \tau}
+\frac{1}{\bar{\nu}(\eta)}\frac{\partial [\bar{\nu}(\eta) v_z]}{\partial \eta }=
\tau\bar{D}_\Omega(\rho)\big{[}\frac{\partial v_r}{\partial \Theta}
-2v_\theta\big{]}
-\frac{1}{\rho}\big{[}\frac{\partial v_\theta }{\partial \Theta}
+v_r\big{]}
-\frac{\partial v_r}{\partial \rho}
-[\bar{D}_\Omega(\rho)+\bar{D}_N(\rho)]v_r,
\label{62}
\end{equation}
\begin{equation}
\frac{\partial b_\theta}{\partial \tau}
+\frac{\partial v_z}{\partial \eta }=
\bar{D}_\Omega(\rho)\big{[}\frac{\partial v_r}{\partial \Theta}
-2v_\theta\big{]}-\frac{3}{2}b_r
+\frac{1}{\bar{S}(\rho)}\frac{\partial v_\theta }{\partial \eta}
-\frac{\partial v_r}{\partial \rho}
-[\bar{D}_\Omega(\rho)+\frac{1}{2}\bar{D}_{B\theta}(\rho)]v_r.
\label{63}
\end{equation}

{\it The  poloidal magnetic field in the IC-regime.}
\begin{equation}
 \frac{\partial b_r}{\partial \tau}=
\frac{1}{\rho}\frac{\partial v_r}{\partial \Theta }
+\frac{1}{\bar{S}(\rho)}
\frac{\partial v_r}{\partial \eta },
\label{64}
\end{equation}
\begin{equation}
 \frac{\partial b_z}{\partial \tau}=
  \tau
\frac{\bar{D}_\Omega(\rho)}{\bar{S}(\rho)}
\big{[}\frac{\partial v_r}{\partial \Theta}
-2v_\theta\big{]}
+\frac{1}{\rho}
\frac{\partial v_z}{\partial \Theta}
- \frac{1}{\bar{S}(\rho)}\big{[}\frac{1}{\rho}\frac{\partial v_\theta}{\partial \Theta}
+\frac{\partial v_r}{\partial \rho }
-\big{(}\bar{D}_\Omega(\rho)+\frac{1}{2}\bar{D}_{Bz}(\rho)
+\frac{1}{\rho}\big{)}v_r\big{]}.
\label{65}
\end{equation}
%
Again, if the divergence-free condition,  $\nabla \cdot {\bf{B}}'=0$, is satisfied at the initial instant, it will be fulfilled for all subsequent times. Additionally, Eq. (\ref{63}) for the perturbed toroidal magnetic field was derived by employing the divergence-free condition of the magnetic field.
The system of Eqs. (\ref{59})- (\ref{65}) is subject to the vanishing boundary conditions at infinity for the perturbed density and magnetic field.


\subsection{Normal-mode stability for the IC-regime.}
In order to describe the IC -regime, we start again by representing the perturbations in the normal-mode form for any perturbed variable $f$, with the complex eigenvalue $\lambda$  and the azimuthal wavenumber $m=0,\pm1,\pm2,...$
\begin{equation}
f(\rho,\Theta,\eta,\tau)=\exp[-i \lambda\tau+im\Theta ] \hat{f}(\rho,\eta)+c.c.\,\,\,\,
\label{66}
\end{equation}
Substituting that ansatz into the governing equations reveals that the spectral properties of the IC-regime are identical to those that characterize the MS-regime. Indeed, the subsystems (\ref{59}) -(\ref{60}) and (\ref{61}) -(\ref{63}) for the IC and MS modes of the IC-regime are described by (\ref{52})- (\ref{54}) and  (\ref{46})-(\ref{50}) for MS-regime in Section 3. However, as will be seen in the next sub-section, while in the IC regime the MS modes drive the IC waves to non-exponential growth, the opposite happens in the MS regime. As in Section 3, the resulting equations which describe the temporal growth of the perturbations don't contain  radial derivatives, the normal-mode form (66) is used without assuming an exponential form in the radial direction with the corresponding characteristic scale  $\varkappa_r\sim\epsilon^0$.

\subsection{Algebraic temporal growth of perturbations in the IC-regime.}

Assuming that the MS modes are non-resonantly driven by IC- modes (that is determined by (\ref{52}) - (\ref{54})) and substituting (\ref{67}) into in the right-hand sides of  (\ref{61}) - (\ref{63}) for the MS mode and  (\ref{64})- (\ref{65}) for the poloidal magnetic fields $\{b_r, b_z\}$, yields (keeping for brevity only linear in $\tau$  terms in the right-hand sides (\ref{59}) - (\ref{65})):
\begin{equation}
\{v_z,\nu, b_r, b_\theta, b_z\}=\tau\{\hat{v}_z^{(1)},\hat{\nu}^{(1)}, \hat{b}_r^{(1)}, \hat{b}_\theta^{(1)},\hat{b}_z^{(1)})\}
(\rho,\eta)\exp(-i\lambda_{IC}\tau+im\Theta)+c.c.\,\,\,\,\,\,
\label{67}
\end{equation}
Assuming again non-resonant conditions and using  $\lambda_{IC}=\pm1$, yields after cumbersome calculations the following partial solution that is driven by the IC mode in the right-hand sides (\ref{59}) - (\ref{65}):
$$
\hat{b}_r^{(1)}=0,\,\,\,\,\,
\,
\hat{b}_\theta^{(1)}=\frac{\bar{D}_\Omega(\rho)\bar{\beta}_\theta(\rho)}
{i\lambda_{IC}[\bar{\beta}_\theta(\rho)\bar{\nu}(\eta)+1]}
\int(im\hat{v}_r-2\hat{v}_\theta)d\bar{\nu}(\eta),
\,\,\,\,\,\,\,
\hat{b}_z^{(1)}=-\frac{\bar{D}_\Omega(\rho)}
{i\lambda_{IC}\bar{S}(\rho)}
(im\hat{v}_r-2\hat{v}_\theta),\,\,\,\,\,\,\,\,\,\,\,\,\,\,\,\,\,\,\,\,\,\,\,\,\,\,\,\,\,\,\,\,\,\,\,\,\,\,\,\,\,\,\,\,\,\,\,
\,\,\,\,\,\,\,\,\,\,\,\,\,\,\,\,\,\,\,\,\,\,\,\,\,\,\,\,\,\,\,\,\,\,\,\,\,\,\,\,\,\,
\,\,\,\,\,\,\,\,\,\,\,\,\,\,\,\,\,\,\,\,\,\,\,\,\,\,\,\,\,\,\,\,\,\,\,\,\,\,\,\,\,\,
$$
\begin{equation}
\hat{v}_z^{(1)}=\bar{D}_\Omega(\rho)\int(im\hat{v}_r-2\hat{v}_\theta)d\eta
+i\lambda_{IC}\int \hat{b}_\theta^{(1)}d\eta,\,\,
\hat{\nu}^{(1)}=\hat{b}_\theta^{(1)}+
\frac{1}{i\lambda_{IC}}\frac{1}{\bar{\nu}(\eta)}
\frac{d\bar{\nu}(\eta)}{d\eta}\hat{v}_z^{(1)}.\,\,\,\,\,\,
\label{68}
\end{equation}
Expressions (\ref{68}) determine the perturbed magnetic field, density,  and axial velocity,
 through the radial and toroidal velocities, $(im\hat{v}_r$, $\hat{v}_\theta)$, which are the eigenfunctions for the IC mode introduced in (\ref{54}),  arbitrary functions, $F(\rho)$  and   $G(\rho)$,  should be additionally specified in order to  satisfy the corresponding boundary conditions.

\section{MIXED IC-MS REGIME WITH HIGH AZIMUTHAL AND RADIAL WAVENUMBERS.}
\subsection{ General magnetohydrodynamic system.}
Within each of the two regimes discussed above, namely the IC and MS regimes, the two modes of wave propagation (IC and MS waves) are practically decoupled, as they may exist independently of each other. In addition, within each of the regimes one of the modes may also drive the other. To avoid such mode decoupling, the effect of large values  $k_r, k_\theta\sim k_z\sim\epsilon$   is examined in accordance with the principle of  least possible degeneracy of the problem (where  $\bold{k}=\{k_r,k_\theta,k_z\}$ is the wave vector,  $k_r=2\pi/\varkappa_r$,  $k_\theta=2\pi/\varkappa_\theta$,  $k_z=2\pi/\varkappa_z$,  see the last row of  Table 2). As is subsequently shown for perturbations of such scale lengths, both modes are coupled, and the disc exhibits a strong instability which is absent within IC and MS regimes discussed above.
As the radial as well as the azimuthal scale lengths of the perturbations are of order  $\epsilon$, the perturbations are represented in the following normal form  in the axial and the radial variables:
\begin{equation}
f(\rho,\Theta,\eta,\tau)=
\exp[i(k_\rho\rho+m_\Theta\Theta+k_\eta\eta)]\hat{f}(\rho,\eta,\tau)+c.c.,
\label{69}
\end{equation}
where to  leading order in  $\epsilon$,  $\hat{f}(\rho,\eta,\tau)$ parametrically depends on $\rho$, and is independent of $\eta$  (i.e.  the disc structure in $\eta$ may be ignored),
$$
k_\rho\cong \epsilon k_r\sim \epsilon^0,\,\,\,
k_\Theta \cong \epsilon k_\theta \equiv  m_\Theta/\rho\sim \epsilon^0
,\,\,\,(m_\Theta\cong \epsilon m\sim \epsilon^0),\,\,\,\,
k_\eta\cong\epsilon k_z\sim \epsilon^0.
\,\,\,\,\,\,\,\,\,\,\,\,\,\,\,\,\,\,\,\,\,\,\,\,\,\,\,\,\,\,\,\,\,\,\,
\,\,\,\,\,\,\,\,\,\,\,\,\,\,\,\,\,\,\,\,\,\,\,\,\,\,\,\,\,\,\,\,\,\,\,\,\,\,\,\,\,\,
\,\,\,\,\,\,\,\,\,\,\,\,\,\,\,\,\,\,\,\,\,\,\,\,\,\,\,\,\,\,\,\,\,\,\,\,\,\,\,\,\,\,
\,\,\,\,\,\,\,\,\,\,\,\,\,\,\,\,\,\,\,\,\,\,\,\,\,\,\,\,\,\,\,\,\,\,\,\,\,\,\,\,\,\,
\,\,\,\,\,\,\,\,\,\,\,\,\,\,\,\,\,\,\,\,\,\,\,\,\,\,\,\,\,\,\,\,\,\,\,\,\,\,\,\,\,\,
$$
Employing now the expansions described in the last row of Table 2, yields:
\begin{equation}
 \frac{\partial \hat{v}_r}{\partial \tau}
-2\hat{v}_\theta=
-i\Delta k_\rho (\tau)\hat{\nu}
+\frac{i}{k_\Theta}\frac{1}{\bar{\beta}_\theta}
\{[\Delta k_\rho^2(\tau)+k_\Theta^2]\hat{b}_r
+\Delta k_\rho(\tau)k_\eta\hat{b}_z\}
,
\label{70}
\end{equation}
\begin{equation}
\frac{\partial \hat{v}_\theta }{\partial \tau}
+\frac{1}{2} \hat{v}_r=
-i k_\Theta \hat{\nu},
\label{71}
\end{equation}
\begin{equation}
\frac{\partial \hat{v}_z}{\partial \tau}=
- i k_z \hat{\nu}
+\frac{i}{k_\Theta}\frac{1}{\bar{\beta}_\theta}
[(k_\Theta^2+k_\eta^2)\hat{b}_z
+\Delta k_\rho(\tau)k_\eta\hat{b}_r],
\label{72}
\end{equation}
\begin{equation}
\frac{\partial \hat{\nu}}{\partial \tau}
=-ik_z\hat{v}_z
-i\Delta k_\rho(\tau)\hat{v}_r
-i k_\Theta\hat{v}_\theta,
\label{73}
\end{equation}
\begin{equation}
\frac{\partial \hat{b}_r}{\partial \tau}=i k_\Theta\hat{v}_r,
\label{74}
\end{equation}
\begin{equation}
 \frac{\partial \hat{b}_\theta}{\partial \tau}=
-i k_z\hat{v}_z-i\Delta k_\rho(\tau)\hat{v}_r-\frac{3}{2}\hat{b}_r,
\label{75}
\end{equation}
\begin{equation}
\frac{\partial \hat{b}_z}{\partial \tau}=i k_\Theta\hat{v}_z,
\label{76}
\end{equation}
where $\Delta k_\rho(\tau)= k_\rho-\tau\rho\bar{D}_\Omega(\rho)k_\Theta(\rho)$.
 Equations (\ref{72})-(\ref{78}) imply that
\begin{equation}
\frac{\partial \nabla\cdot\bold{\hat{b}}}{\partial \tau}=0.
\label{77}
\end{equation}
Hence, the divergence-free condition is satisfied at all times by choosing a proper initial condition for the magnetic field.

Assuming now large plasma beta, and applying the asymptotic principle of least possible degeneracy of the problem in the limit  $\bar{\beta}_\theta\gg1$ leads to the following estimations:
\begin{equation}
k_\rho\sim k_\Theta\sim k_\eta\sim\Delta k_\rho(\tau)\sim\sqrt{\bar{\beta}}_\theta,\,\,\,\,\,
\hat{\nu}\sim\bar{\beta}_\theta^{-1}\hat{b}_r\sim\bar{\beta}_\theta^{-1}\hat{b}_z.\,\,\,\,\,
\label{78}
\end{equation}
Choosing now for further convenience  $k_\Theta$ as the basic parameter instead of   $\bar{\beta}_\theta$,  the  new scaled variables of order of unity in $k_\Theta\sim\bar{\beta}_\theta^2$   are introduced
\begin{equation}
\tilde{k}_\rho=\frac{k_\rho}{k_\Theta},\,\,\,\,\,\
\tilde{k}_\Theta=\frac{k_\Theta}{k_\Theta}\equiv 1,\,\,\,\,\,\
\tilde{k}_\eta=\frac{k_\eta}{k_\Theta},\,\,\,\,\,\
\Delta \tilde{k}_\rho(\tau)=\frac{\Delta k_\rho(\tau)}{k_\Theta},\,\,\,\,\,\
\Delta \tilde{k}(\tau)=\frac{\Delta k(\tau)}{k_\Theta},\,\,\,\,\,\
\tilde{\beta}_\theta=\frac{\bar{\beta}_\theta}{k_\Theta^2},
\label{79}
\end{equation}
\begin{equation}
\tilde{\nu}_\rho=\hat{\nu}k_\Theta^2,\,\,\,\,\,\
\tilde{b}_r=\hat{b}_r,\,\,\,\,\,\
\tilde{b}_\Theta=\hat{b}_\Theta,\,\,\,\,\,\
\tilde{b}_z=\hat{b}_z,\,\,\,\,\,\
\tilde{v}_r=k_\Theta\hat{v}_r,\,\,\,\,\,\
\tilde{v}_\Theta=k_\Theta\hat{v}_\Theta,\,\,\,\,\,\
\tilde{v}_z=k_\Theta\hat{v}_z,\,\,\,\,\,\
\label{80}
\end{equation}
where
\begin{equation}
\Delta \tilde{k}^2(\tau)=
\Delta \tilde{k}_\rho^2(\tau)+1+\tilde{k}_\eta^2,\,\,\,\,\,\
\Delta \tilde{k}_\rho(\tau)=\tilde{k}_\rho-\rho\bar{D}_\Omega(\rho) \tau
\equiv\tilde{k}_\rho+\frac{3}{2}\tau.
\label{81}
\end{equation}
This yields the following equations with coefficients of  order of unity in the plasma beta parameter:
\begin{equation}
 \frac{\partial i\tilde{v}_r}{\partial \tau}
-2i\tilde{v}_\theta=
\Delta \tilde{k}_\rho (\tau)\tilde{\nu}
-\frac{1}{\tilde{\beta}_\theta}
\{[\Delta \tilde{k}_\rho^2(\tau)+1]\tilde{b}_r
+\Delta \tilde{k}_\rho(\tau)\tilde{k}_\eta\tilde{b}_z\}
,
\label{82}
\end{equation}
\begin{equation}
\frac{\partial i \tilde{v}_\theta }{\partial \tau}
+\frac{1}{2} i\tilde{v}_r=
\tilde{\nu},
\label{83}
\end{equation}
\begin{equation}
\frac{\partial i \tilde{v}_z}{\partial \tau}=
\tilde{k}_\eta \tilde{\nu}
-\frac{1}{\tilde{\beta}_\theta}
[(\tilde{k}_\eta^2+1)\tilde{b}_z
+\Delta \tilde{k}_\rho(\tau)\tilde{k}_\eta\tilde{b}_r],
\label{84}
\end{equation}
\begin{equation}
i\Delta \tilde{k}_\rho(\tau)\tilde{v}_r
+i \tilde{v}_\theta+i\tilde{k}_\eta\tilde{v}_z=0
,
\label{85}
\end{equation}
\begin{equation}
\frac{\partial \tilde{b}_r}{\partial \tau}=i \tilde{v}_r,
\label{86}
\end{equation}
\begin{equation}
 \frac{\partial \tilde{b}_\theta}{\partial \tau}=
-i \tilde{k}_\eta\tilde{v}_z-i\Delta \tilde{k}_\rho(\tau)\tilde{v}_r-\frac{3}{2}\tilde{b}_r,
\label{87}
\end{equation}
\begin{equation}
\frac{\partial \tilde{b}_z}{\partial \tau}=i\tilde{v}_z.
\label{88}
\end{equation}
Here the parametric dependence on  $\rho$  is omitted for brevity; the time derivative of the perturbed density, $\tilde{\nu}$,   in  (\ref{85}) is neglected due to its smallness at large plasma beta   $\bar{\beta}_\theta\gg 1$, while the steady-state (\ref{85}) is reduced to the divergence-free condition for the plasma velocity. It is interesting to note that condition (\ref{85}), which is tantamount to the Boussinesq approximation, is  not assumed, but rigorously obtained from the expansion scheme in small $\epsilon$ and  $1/\bar{\beta}_\theta\gg 1$. It can be directly verified  using (\ref{86}) -(\ref{88}) that the divergence-free condition  (\ref{85}) for the velocity  $\tilde{\bold{v}}$  is satisfied identically due to the divergence-free condition for the magnetic field, rewritten as
\begin{equation}
i\Delta \tilde{k}_\rho(\tau)\tilde{b}_r
+i \tilde{b}_\theta+i\tilde{k}_\eta\tilde{b}_z=0.
\label{89}
\end{equation}

Equations (\ref{82}) -(\ref{88})   may be now  reduced to the following coupled system of equations for the poloidal magnetic field and the  density perturbations:
\begin{equation}
\frac{\partial^2 \tilde{b}_r}{\partial \tau^2}=
-2\tilde{k}_\eta \frac{\partial \tilde{b}_z}{\partial \tau}
-2\Delta \tilde{k}_\rho (\tau)\frac{\partial \tilde{b}_r}{\partial \tau}
+\Delta \tilde{k}_\rho (\tau)\tilde{\nu}
-\frac{1}{\tilde{\beta}_\theta}
\{[\Delta \tilde{k}_\rho ^2(\tau)+1]\tilde{b}_r
+\Delta \tilde{k}_\rho (\tau)\tilde{k}_\eta \tilde{b}_z
\}
,
\label{90}
\end{equation}
\begin{equation}
\frac{\partial^2 \tilde{b}_z}{\partial \tau^2}=
\tilde{k}_\eta \tilde{\nu}
-\frac{1}{\tilde{\beta}_\theta}
[(1+\tilde{k}_\eta ^2)\tilde{b}_z
+\Delta \tilde{k}_\rho(\tau)\tilde{k}_\eta \tilde{b}_r],
\label{91}
\end{equation}
\begin{equation}
\tilde{\nu}=\frac{1}{\Delta \tilde{k}^2(\tau)}
[2\Delta \tilde{k}_\rho^2(\tau)-1]
\frac{\partial \tilde{b}_r}{\partial \tau}
+2\frac{\tilde{k}_\eta\Delta \tilde{k}_\rho(\tau)}{\Delta \tilde{k}^2(\tau)}
\frac{\partial \tilde{b}_z}{\partial \tau}
+\frac{1}{\tilde{\beta}_\theta}\Delta \tilde{k}_\rho(\tau)\tilde{b}_r
+\frac{1}{\tilde{\beta}_\theta}\tilde{k}_\eta\tilde{b}_z
.
\label{92}
\end{equation}
Substituting $\tilde{\nu}$  from (\ref{92}) into (\ref{90})-(\ref{91}) finally yields a system of two coupled second order differential equations:
\begin{equation}
\frac{\partial^2 \tilde{b}_r}{\partial \tau^2}
+(3+2\tilde{k}_\eta^2)
\frac{\Delta \tilde{k}_\rho(\tau)}{\Delta \tilde{k}^2(\tau)}
\frac{\partial \tilde{b}_r}{\partial \tau}
+2 \tilde{k}_\eta
\frac{1+\tilde{k}_\eta^2}{\Delta \tilde{k}^2(\tau)}
\frac{\partial \tilde{b}_z}{\partial \tau}
+\frac{1}{\tilde{\beta}_\theta}\tilde{b}_r=0
,
\label{93}
\end{equation}
\begin{equation}
\frac{\partial^2 \tilde{b}_z}{\partial \tau^2}
-\tilde{k}_\eta^2
\frac{2\Delta \tilde{k}_\rho^2(\tau)-1}{\Delta \tilde{k}^2(\tau)}
\frac{\partial \tilde{b}_r}{\partial \tau}
-2 \tilde{k}_\eta^2
\frac{\Delta \tilde{k}_\rho(\tau)}{\Delta \tilde{k}^2(\tau)}
\frac{\partial \tilde{b}_z}{\partial \tau}
+\frac{1}{\tilde{\beta}_\theta}\tilde{b}_z=0.
\label{94}
\end{equation}
After $\tilde{b}_r$  and $\tilde{b}_z$   are found from (\ref{93}) - (\ref{94}), the density   is determined by (\ref{92}), and the rest of the variables  $\tilde{v}_r$, $\tilde{v}_z$    and $\tilde{v}_\theta$, $\tilde{b}_\theta$  can be determined from the (\ref{85})-(\ref{86}) and (\ref{88})- (\ref{89}), respectively:
\begin{equation}
i\tilde{v}_r=\frac{\partial \tilde{b}_r}{\partial \tau},\,\,\,\,
i\tilde{v}_z=\frac{\partial \tilde{b}_z}{\partial \tau},\,\,\,\,
i\tilde{v}_\theta=-\Delta \tilde{k}_\rho(\tau)
\frac{\partial \tilde{b}_r}{\partial \tau}-
\tilde{k}_\eta\frac{\partial \tilde{b}_z}{\partial \tau},\,\,\,\,
i\tilde{b}_\theta=-\Delta \tilde{k}_\rho(\tau) \tilde{b}_r-
\tilde{k}_\eta  \tilde{b}_z.\,\,\,\,
\label{95}
\end{equation}

Numerical solution of (\ref{92})- (\ref{95}) for amplitudes of the perturbed magnetic field, velocity and density vs the orbital time  $\tau$ in the mixed IC- MS (oscillatory) regime of instability are presented in Figs. 1-4 for typical values of the scaled (with the azimuthal wave number) parameters,  $\tilde{k}_\rho$,  $\tilde{k}_\eta$,  $\tilde{\beta}_\theta$. It is noted that without loss of generality  $\tilde{k}_\eta$  may be assumed to be positive since changing  its sign is equivalent to sign changing of either $\tilde{b}_r$  or  $\tilde{b}_z$. Figures 1-4 demonstrate that the axial magnetic component  $\tilde{b}_z$ dominates the radial one  $\tilde{b}_r$ for all values of the parameters.  Moreover,
a significant amplitude growth of the axial magnetic field $\tilde{b}_z$   may be achieved after tens of orbital times.
It should be noted that the results obtained for the mixed IC-MS regime of instability are quite similar to those in \cite{Balbus and Hawley 1992}.
Such a growth of  $\tilde{b}_z$ may  have a non-monotonic character which is evident by its decrease with further increasing of time, i.e. the instability may be of transient nature.
%
Figures 1-4 also exhibit strong dependence of the magnetic field, velocity and density on the values of the parameters.
It is  noted that increasing $\tilde{\beta}_\theta$  and  $\tilde{k}_\eta$  has a dramatic effect on amplification of the initial perturbations. Thus, the pure toroidal magnetic field configuration is susceptible to high beta, high small axial wave length, and low azimuthal wave numbers instabilities. In particular, a significant amplitude growth of the perturbations by  factors $\sim 10^4\div10^6$  may be achieved after tens of orbital times (see Figures 4 and 5).

\begin{figure*}
\includegraphics[width=160mm]{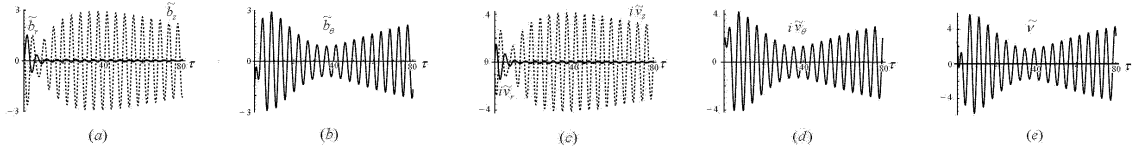}
  \caption{
  Amplitudes of perturbed magnetic field, velocity and density  vs orbital time  in the mixed IC- MS  regime of instability for typical parameters  $\tilde{k}_\rho=0.5$,
 $\tilde{k}_\eta=1.5$,  $\tilde{\beta}_\theta=0.5$
and partial initial data  $\tilde{b}_r(0)=0$,  $i\tilde{v}_r(0)\equiv\partial\tilde{b}_r/\partial\tau(0)=0$,
 $\tilde{b}_z(0)=1$,  $i\tilde{v}_z(0)\equiv\partial\tilde{b}_z/\partial\tau(0)=-1$.
(a) poloidal magnetic fields, $\tilde{b}_r$ (solid line) and  $\tilde{b}_z$  (dashed line); (b) toroidal magnetic field,  $\tilde{b}_\theta$;
(c) poloidal velocities, $i\tilde{v}_r$ (solid line) and   $i\tilde{v}_z$  (dashed line); (d) toroidal velocity,   $\tilde{v}_\theta$; (e) density,   $\tilde{\nu}$.
  }
\label{Fig. 1}
\end{figure*}

\begin{figure*}
\includegraphics[width=160mm]{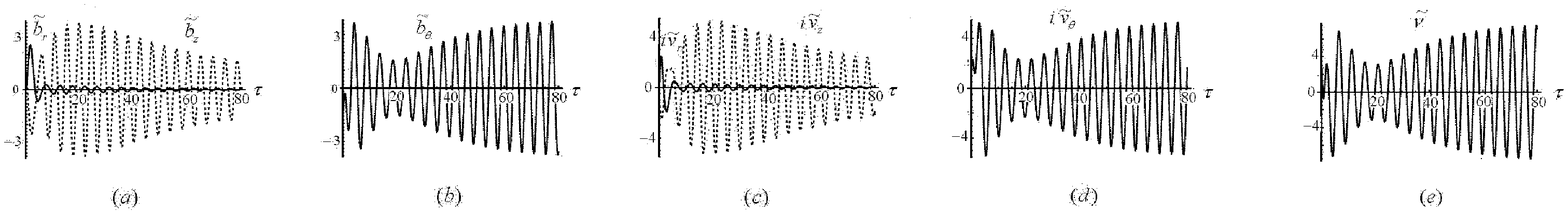}
  \caption{
 The same as in Fig. 1 for  $\tilde{k}_\rho=-0.5$,
 $\tilde{k}_\eta=1.5$,  $\tilde{\beta}_\theta=0.5$
.
  }
\label{Fig. 2}
\end{figure*}

\begin{figure*}
\includegraphics[width=160mm]{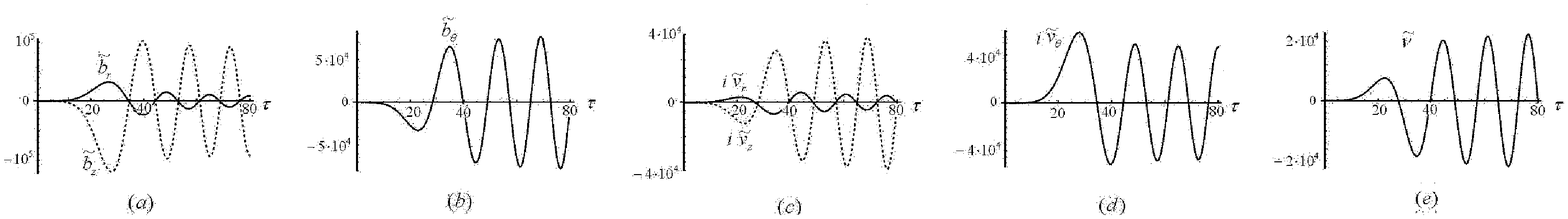}
  \caption{
 The same as in Fig. 1 for  $\tilde{k}_\rho=3.5$,
 $\tilde{k}_\eta=12$,  $\tilde{\beta}_\theta=3.5$
.
  }
\label{Fig. 3}
\end{figure*}

\begin{figure*}
\includegraphics[width=160mm]{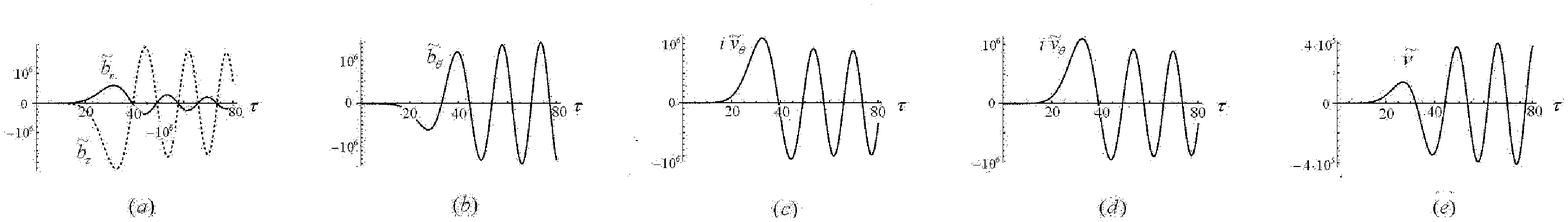}
  \caption{
 The same as in Fig. 1 for  $\tilde{k}_\rho=-3.5$,
 $\tilde{k}_\eta=12$,  $\tilde{\beta}_\theta=3.5$
.
  }
\label{Fig. 4}
\end{figure*}

\section{THE PURE HYDRODYNAMIC LIMIT.}

To elucidate the role of the magnetic fields in the perturbations growth, consider the pure hydrodynamic limit by setting both equilibrium and perturbed magnetic fields to zero, while maintaining the same assumptions and notations in all above three regimes.
In that case the regimes of instability may be classified by a quite similar way as for  the general MHD case. There are three following regimes: the MS regime (degenerated in the pure hydrodynamic limit to the acoustic regime) and the IC regime, which are characterized by algebraic temporal growth of the perturbations, as well as the mixed IC-acoustic regime for which the non-algebraic instability has been established for non-axisymmetric perturbations.   Below, we restrict ourselves  the mixed IC-acoustic regime as the most degenerous one, while the first two regimes are not described here for brevity.
%
%
%
%

%

{\it The mixed IC-acoustic regime.}
In a similar way to the mixed IC-MS-regime the pure hydrodynamic limit is as follows:
\begin{equation}
\frac{\partial \tilde{v}_r}{\partial \tau}-2\tilde{v}_\theta=
-\frac{\Delta \tilde{k}_\rho (\tau)}{\Delta \tilde{k}^2 (\tau)}
( \tilde{v}_r
+2\tilde{v}_\theta)
,\,\,\,\,
\label{96}
\end{equation}
\begin{equation}
\frac{\partial \tilde{v}_\theta}{\partial \tau}+\frac{1}{2}\tilde{v}_r=
-\frac{1}{\Delta \tilde{k}^2 (\tau)}
( \tilde{v}_r
+2\tilde{v}_\theta)
,\,\,\,\,
\label{97}
\end{equation}
\begin{equation}
 \tilde{\nu}=
-\frac{i}{\Delta \tilde{k}^2 (\tau)}
( \tilde{v}_r
+2\tilde{v}_\theta)
.\,\,\,\,
\label{98}
\end{equation}
Equation (\ref{98}) expresses the divergence-free condition for the perturbed velocity. The results of the numerical solutions of Eqs. (\ref{96})-(\ref{98}) are presented in Figs. 5-8 for the same values of the parameters as for the  MHD system in Figs. 1-4, respectively.

\begin{figure*}
\includegraphics[width=120mm]{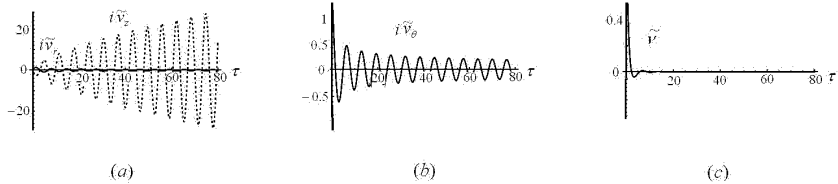}
  \caption{
Amplitudes of perturbed velocity and density  vs orbital time  in the mixed IC- MS pure hydrodynamic regime of instability for typical parameters  $\tilde{k}_\rho=0.5$,
 $\tilde{k}_\eta=1.5$,  $\tilde{\beta}_\theta=0.5$
and partial initial data  $i\tilde{v}_r(0)=0$, $i\tilde{v}_z(0)=-1$  (or, equivalently,  $\tilde{v}_\theta(0)=\tilde{k}_\eta$).
 (a) poloidal velocities, $i\tilde{v}_r$ (solid line) and  $i\tilde{v}_z$   (dashed line); (b) toroidal velocity,  $\tilde{v}_\theta$; (c) density,   $\tilde{\nu}$.
  }
\label{Fig. 5}
\end{figure*}

\begin{figure*}
\includegraphics[width=120mm]{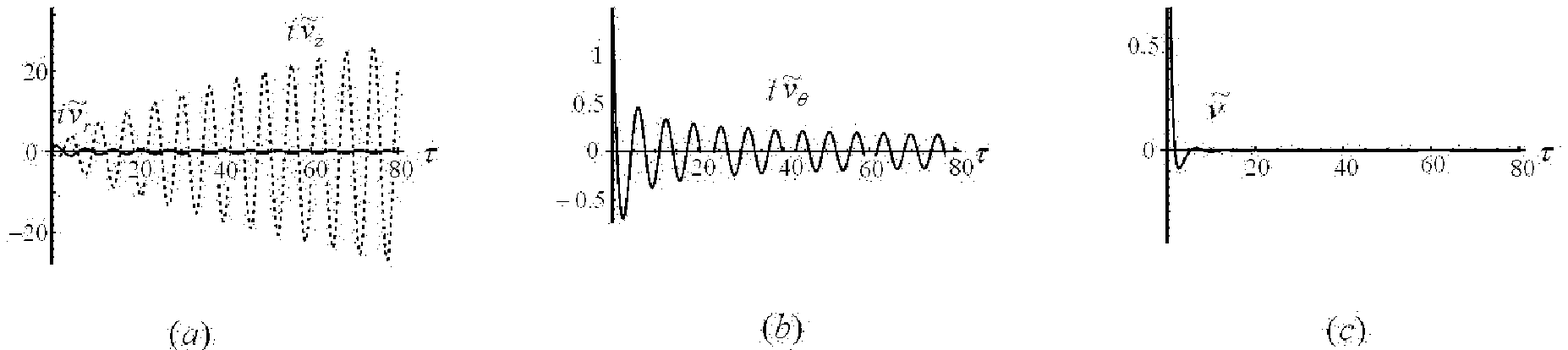}
  \caption{
The same as in Fig. 5 for  $\tilde{k}_\rho=-0.5$,
 $\tilde{k}_\eta=1.5$,  $\tilde{\beta}_\theta=0.5$
 .
  }
\label{Fig. 6}
\end{figure*}

\begin{figure*}
\includegraphics[width=120mm]{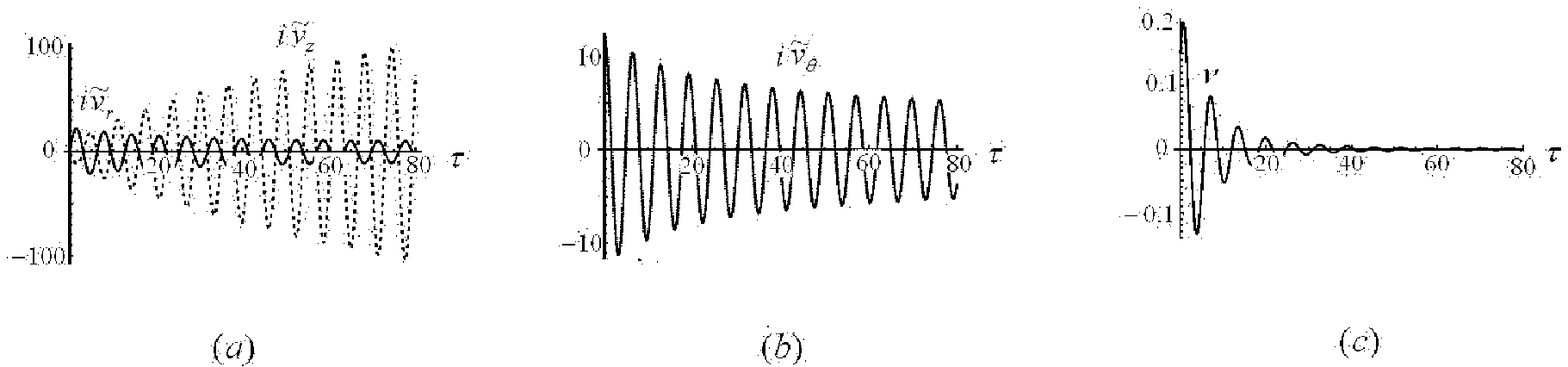}
  \caption{
The same as in Fig. 5 for  $\tilde{k}_\rho=3.5$,
 $\tilde{k}_\eta=12$,  $\tilde{\beta}_\theta=3.5$
  .
  }
\label{Fig. 7}
\end{figure*}

\begin{figure*}
\includegraphics[width=120mm]{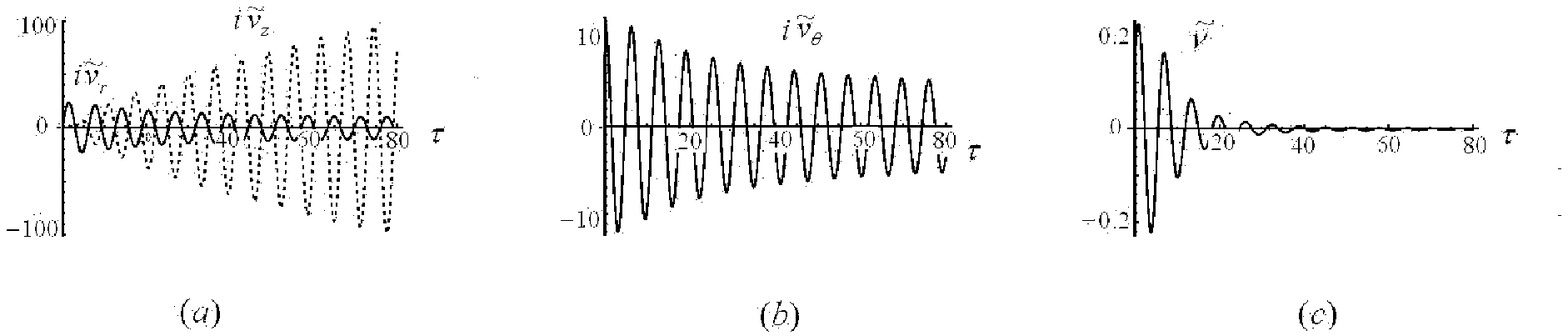}
  \caption{
The same as in Fig. 5 for  $\tilde{k}_\rho=-3.5$,
 $\tilde{k}_\eta=12$,  $\tilde{\beta}_\theta=3.5$
   .
  }
\label{Fig. 8}
\end{figure*}

Comparing  equations for the MS- and IC-regimes
in the general MHD systems
with their counterparts
in the pure hydrodynamic limits,
it  can be shown  that at large plasma beta the magnetic field has no significant influence on the linear temporal growth of the in-plane velocities. In contrast, the behavior of the pure hydrodynamic and MHD systems is quite different for the mixed IC- MS regime, where the presence of magnetic fields results in a dramatic increase of the perturbation  growth rates.
It should be noted that the pure hydrodynamic limit presented in this section generalizes previous hydrodynamical results such as \cite{ Umurhan et al. 2006}, \cite{Rebusco et al. 2009} and SML2011. The latter focused only on those perturbations which belong under the current classification to the IC regime (i.e. are scaled according to the fourth row in Table 2). Indeed, typical amplification of such perturbations does not exceed 1-2 orders of magnitude after about 100 rotating times [see \cite{ Umurhan et al. 2006} and \cite{Rebusco et al. 2009}]. This is in contrast to the much higher amplification of perturbations that belong to the mixed regime under the influence of a toroidal magnetic field, which may exceed six orders of magnitude (compare Figs. 4 and 8).

\section{Summary and discussion.}
The present study concerns the stability of thin density-stratified vertically-isothermal Keplerian discs in a toroidally-dominated magnetic field. The problem is treated by asymptotic expansions in small aspect ratio of the disc. The
vertically-isothermal thin
disc approximation is found as the most suitable model to stability study of Keplerian discs due to its relative simplicity and adequacy to true disc geometry. The model may be effectively combined with the local approximations (as it is done here for the mixed IC-MS regime). The toroidally-dominated magnetic configurations are found to be stable with respect to normal-mode perturbations. Instead, three non-exponential regimes of temporal growth of perturbations governed by the rotation shear have been identified depending on the initial level of perturbed velocities. First two regimes (magneto-sonic (MS) and inertia-coriolis (IC) regimes are driven by MS- and IC-modes, and decoupled into in-plane and normal-plane modes with small in-plane components of velocity compared with axial velocity component and contrariwise, respectively.  The MS- and IC- regimes support the perturbations with algebraically growing amplitudes for both axisymmetric and non-axisymmetric modes. The third, mixed IC-MS regime is pure non-axisymmetric, operates on high radial and azimuthal wavenumbers, and corresponds to the least possible degeneration of the problem with comparable in-plane and axial perturbed velocities. This regime most likely corresponds to maximal growing perturbations, which demonstrate significant growth rates with a growth time measured in tens of orbital periods. The corresponding time-dependences of the magnetic fields are qualitatively similar to that in \cite{Balbus and Hawley 1992}. Moreover,  although the compressibility effects are taken into account consequently through the current analysis, the divergence-free condition for the Bousinesq approximation is valid in both models: as a'priori accepted in \cite{Balbus and Hawley 1992} modeling, or as the corollary of the high plasma beta approximation in the present analysis. In the first two regimes of instability the compressible MS mode plays a principal role either as the driver of the growth or the driven growing mode, while the mixed IC-MS regime is described by the Bousinesq approximation for incompressible fluid that is obtained as a natural limit of the expansion scheme. Additionally, as distinct from IC and MS regimes, behavior of the pure hydrodynamic and MHD systems for the mixed IC- MS regime, the presence of magnetic fields may drastically increase growth rates of perturbations.

\section*{Acknowledgments}
The current work has been supported by grant no. 180/10 of the Israel Science Foundation.


\bsp
\label{lastpage}
\end{document}